\documentclass[pre,showpacs,floats,twocolumn,floatfix,superscriptaddress]{revtex4-1}

\usepackage{graphics,graphicx,dcolumn,bm,fleqn,epic,eepic,float}
\usepackage{amssymb,amsmath,multirow,rotate,rotating,color}
\usepackage[utf8]{inputenc}

\newcommand{\figref}[1]{Fig.~\ref{fig:#1}}
\newcommand{\eqnref}[1]{Eq.~(\ref{eq:#1})}

\newcommand{\vectornorm}[1]{\left|\left|#1\right|\right|}
\renewcommand{\vec}[1]{\mathbf{#1}}
\newcommand{\uvec}[1]{\hat{\vec{#1}}}
\renewcommand{\tensor}[1]{\mathbf{#1}}

\newcommand{\figwidth}{1.0\linewidth}

\begin{document}

\title{Effects of nanoparticles and surfactant on droplets in shear flow}

\author{Stefan Frijters}
\email{s.c.j.frijters@tue.nl}
\affiliation{Department of Applied Physics, Eindhoven University of Technology, Den Dolech 2, NL-5600MB Eindhoven, The Netherlands}

\author{Florian G\"unther}
\email{f.s.guenther@tue.nl}
\affiliation{Department of Applied Physics, Eindhoven University of Technology, Den Dolech 2, NL-5600MB Eindhoven, The Netherlands}

\author{Jens Harting} 
\email{j.harting@tue.nl}
\affiliation{Department of Applied Physics, Eindhoven University of Technology, Den Dolech 2, NL-5600MB Eindhoven, The Netherlands}
\affiliation{Institute for Computational Physics, University of Stuttgart, Pfaffenwaldring 27, D-70569 Stuttgart, Germany} 

\date{\today}

\begin{abstract}
We present three-dimensional numerical simulations, employing the well-established lattice Boltzmann method, and investigate similarities and differences between surfactants and nanoparticles as additives at a fluid-fluid interface. We report on their respective effects on the surface tension of such an interface. Next, we subject a fluid droplet to shear and explore the deformation properties of the droplet, its inclination angle relative to the shear flow, the dynamics of the particles at the interface, and the possibility of breakup. Particles are seen not to affect the surface tension of the interface, although they do change the overall interfacial free energy. The particles do not remain homogeneously distributed over the interface, but form clusters in preferred regions that are stable for as long as the shear is applied. However, although the overall structure remains stable, individual nanoparticles roam the droplet interface, with a frequency of revolution that is highest in the middle of the droplet interface, normal to the shear flow, and increases with capillary number. We recover Taylor's law for small deformation of droplets when surfactant or particles are added to the droplet interface. The effect of surfactant is captured in the capillary number, but the inertia of adsorbed massive particles increases deformation at higher capillary number and eventually leads to easier breakup of the droplet.
\end{abstract}


\maketitle

\section{Introduction}
\label{sec:introduction}

Stabilizing emulsions by employing nanoparticles is a very attractive tool in the food, cosmetics, oil and medical industries. This method of emulsification complements the traditional use of surfactants -- amphiphilic molecules -- as emulsification agents. Using nanoparticles can have many advantages, such as reduced cost and toxicity and the possibility of tailor-made nanoparticles, which may include useful properties other than being an emulsifier, such as ferromagnetic particles~\cite{ bib:kim-stratford-cates:2010} or Janus particles~\cite{ bib:binks-fletcher:2001}. Although the effects of both emulsifiers can be similar, the underlying physics is very different~\cite{ bib:binks:2002, bib:tcholakova-denkov-lips:2008}.

Amphiphiles are chemical compounds which have both hydrophilic and hydrophobic properties, restricted to specific groups of the molecules. For example, surfactants are characterized by their hydrophilic ``head'' and hydrophobic ``tail(s)''. When they are located at the fluid-fluid interface the possibility exists for both parts of the molecule to reside in their preferred fluid. This makes it energetically favourable for them to accumulate at the interface, with a distinct alignment. This process lowers interfacial tension and prevents the demixing of two immiscible fluids. As such, it gives rise to the possibility of complicated structures, such as micelles and lamellae, gyroid mesophases and the aforementioned emulsion droplets~\cite{ bib:gompper-schick:1994, bib:chen-boghosian-coveney-nekovee:2000, bib:harting-harvey-chin-venturoli-coveney:2005, bib:giupponi-harting-coveney:2006}.

Nanoparticles also find it energetically favourable to adsorp to a fluid-fluid interface, however, this happens for a different reason. Maintaining such an interface requires more energy per unit area than maintaining a particle-fluid interface, and the adsorption of a particle removes the former. Because of the scale of the energy differences involved -- orders of magnitude larger than thermal fluctuations -- this adsorption process tends to be irreversible~\cite{ bib:binks:2002}. In this way, neutrally wetting particles do not affect surface tensions directly, but only change the interfacial free energy. 

When such particles are used to stabilize an emulsion of discrete droplets of one fluid suspended in another, continuous, fluid, the result is known as a ``Pickering emulsion''~\cite{ bib:ramsden:1903, bib:pickering:1907}. The particles in these mixtures block Ostwald ripening, which is one of the main processes leading to drop coarsening in emulsions. Hence, blocking this process allows for a long-term stabilization of such an emulsion. They are also a source of complex rheology due to the irreversible adsorption of the particles as well as interface bridging because of particle monolayers~\cite{ bib:arditty-whitby-binks-schmitt-lealcalderon:2003, bib:arditty-schmitt-giermannskakhan-lealcalderon:2004, bib:binks-clint-whitby:2005}. More recently, the use of nanoparticles has led to the discovery of the ``bicontinuous interfacially jammed emulsion gel'' (commonly referred to as ``bijel''), first predicted by numerical simulations~\cite{ bib:stratford-adhikari-pagonabarraga-desplat-cates:2005} and later confirmed experimentally~\cite{ bib:herzig-white-schofield-poon-clegg:2007, bib:clegg-herzig-schofield-egelhaalf-horozov-binks-cates-poon:2007}. In a bijel, an interface between two continuous fluids (as opposed to having separate droplets of one fluid) is covered and stabilized by particles. The effect of parameters such as fluid:fluid ratio and particle wettability on the final phase a demixing system transforms into has been investigated numerically~\cite{ bib:kim-stratford-adhikari-cates:2008, bib:jansen-harting:2011, bib:aland-lowengrub-voigt:2011, bib:guenther-janoschek-frijters-harting:2012}. 

The differences between the behaviour of amphiphiles and nanoparticles and between their underlying mechanics as described above ensure that many properties of systems including nanoparticles cannot be explained by theories based solely on the physics of amphiphiles. For nanoparticle-stabilized systems, new models have been developed (and verified experimentally), which take into account the features of these systems that have no analogue in surfactant systems, such as the contact angle of the nanoparticles, strong capillary forces between the particles or the pH value and electrolyte concentration of the solvents~\cite{ bib:binks-horozov:2006}. Quantitatively, however, the description of these systems still leaves to be desired.

To properly understand the behaviour of large-scale mixtures with many complex interfaces, such as Pickering emulsions and bijels, one first needs a fundamental understanding of the processes involved on smaller scales. Research was performed to understand in detail how the presence of a nanoparticle~\cite{ bib:degraaf-dijkstra-vanroij:2010} or the collective behaviour of multiple nanoparticles~\cite{ bib:bleibel-dietrich-dominguez-oettel:2011, bib:bleibel-dominguez-oettel-dietrich:2011} affects a flat interface. In this work we investigate the stabilizing effect of amphiphiles or hard spherical nanoparticles on curved interfaces, modeled by a single droplet of a fluid suspended in another fluid. 

Droplets subjected to shear flow display many kinds of interesting behaviour, such as deforming away from a spherical shape, exhibiting an inclination angle with respect to the shear direction and breaking up into smaller droplets (beyond a critical capillary number)~\cite{ bib:janssen-vananroye-vanpuyvelde-moldenaers-anderson:2010}. Nanoparticles adsorped at the droplet interface show an inhomogeneous distribution and a non-trivial motion over the droplet surface. Their presence also affects the deformation and inclination properties of the droplet. We study all these effects in detail in the current article.

Computer simulations are a valuable tool to compare these systems directly, and we choose to employ the lattice Boltzmann (LB) method, which is well-established in the literature (cf.~\cite{ bib:succi:2001}), for our research. The LB method is an alternative to traditional Navier-Stokes solvers, and extensions have been developed to allow for multiple fluids and their interactions~\cite{ bib:shan-chen:1993, bib:shan-chen:1994, bib:orlandini-swift-yeomans:1995, bib:swift-orlandini-osborn-yeomans:1996, bib:dupin-halliday-care:2003, bib:lishchuk-care-halliday:2003}, amphiphiles~\cite{ bib:chen-boghosian-coveney-nekovee:2000, bib:nekovee-coveney-chen-boghosian:2000} and finite-sized particles of arbitrary shape and wettability which can interact with the fluids as well as each other~\cite{ bib:ladd:1994:1, bib:ladd:1994:2, bib:ladd-verberg:2001, bib:joshi-sun:2009, bib:joshi-sun:2010, bib:jansen-harting:2011}.

In section~\ref{sec:sim-method} we introduce the simulation method in detail. Section~\ref{sec:surfacetension} reports and explains our findings on surface tensions in systems of a droplet stabilized by surfactant and nanoparticles. The behaviour of the particles adsorped to the droplet interface when the droplet is subjected to shear is discussed in section \ref{sec:deformation}. This is followed by a an analysis of the effect of nanoparticles and surfactant on the deformation properties and inclination angles of these droplets. The breakup of droplets is then briefly discussed. Finally, conclusions and an outlook are provided in section~\ref{sec:conclusion}.

\section{Simulation method}
\label{sec:sim-method}

\subsection{The lattice Boltzmann method}
\label{ssec:lb}

The lattice Boltzmann method has proven itself to be a very successful tool for modeling fluids in science and engineering~\cite{ bib:chen-doolen:1998, bib:succi:2001, bib:sukop-thorne:2007}. Compared to traditional Navier-Stokes solvers, the method allows an easy implementation of complex boundary conditions and -- due to the high degree of locality of the algorithm -- is well suited for implementation on parallel supercomputers~\cite{ bib:harting-harvey-chin-venturoli-coveney:2005, bib:guenther-janoschek-frijters-harting:2012}.

The method is based on the Boltzmann equation, with its positions $\vec{x}$ discretized in space on a cubic lattice with lattice constant $\Delta x$ and with its time $t$ discretized with a timestep $\Delta t$:
\begin{equation}
  \label{eq:LBG}
  f_i^c(\vec{x} + \vec{c}_i \Delta t , t + \Delta t)=f_i^c(\vec{x},t)+\Omega_i^c(\vec{x},t)
  \mbox{,}
\end{equation}
where $f_i^c(\vec{x},t)$ is the single-particle distribution function for fluid component $c$, being propagated over the lattice with a discrete set of lattice velocities $\vec{c}_i$ and 
\begin{equation}
  \label{eq:BGK_collision_operator}
  \Omega_i^c(\vec{x},t) = -\frac{f_i^c(\vec{x},t)- f_i^\mathrm{eq}(\rho^c(\vec{x},t), \vec{u}^c(\vec{x},t))}{\left( \tau^c / \Delta t \right)}
\end{equation}
is the Bhatnagar-Gross-Krook (BGK) collision operator~\cite{ bib:bhatnagar-gross-krook:1954}. Here, $f_i^\mathrm{eq}(\rho^c,\vec{u}^c)$ is the third-order equilibrium distribution function 
\begin{multline}
  \label{eq:equilibrium-distribution}
  f_i^{\mathrm{eq}}(\rho^c,\vec{u}^c) = \zeta_i \rho^c \cdot \bigg[ 1 + \frac{\vec{c}_i \cdot \vec{u}^c}{c_s^2} + \frac{ \left( \vec{c}_i \cdot \vec{u}^c \right)^2}{2 c_s^4} \\ - \frac{ \left( \vec{u}^c \cdot \vec{u}^c \right) }{2 c_s^2} + \frac{ \left( \vec{c}_i \cdot \vec{u}^c \right)^3}{6 c_s^6} - \frac{ \left( \vec{u}^c \cdot \vec{u}^c \right) \left( \vec{c}_i \cdot \vec{u}^c \right )}{2 c_s^4} \bigg]
  \mbox{,}
\end{multline}
$\tau^c$ is the relaxation time for component $c$ and $\zeta_i$ are the coefficients resulting from the velocity space discretization~\cite{ bib:chen-chen-matthaeus:1992}. We use a three-dimensional lattice and a D3Q19 implementation ($i=1,\ldots,19$), which is to say that $\Delta \vec{x}_i = \vec{c}_i \Delta t$ connect a lattice site with its nearest neighbours and next-nearest neighbours on the lattice. The Navier-Stokes equations can be recovered from \eqnref{LBG}.  The macroscopic densities are given by $\tilde{\rho}^c(\vec{x},t) \equiv \rho^c(\vec{x},t) / \rho^c_0 = \sum_if^c_i(\vec{x},t)$, with $\rho^c_0$ being a reference density for component $c$. For clarity of notation, the tilde is omitted from the densities from now on. The macroscopic velocities are $\vec{u}^c(\vec{x},t) = \sum_if^c_i(\vec{x},t) \vec{c}_i/\rho^c(\vec{x},t)$ in the low Knudsen number and low Mach number limit. The speed of sound on the lattice is 
\begin{equation}
  \label{eq:sos}
  c_S = \frac{1}{\sqrt{3}} \frac{\Delta x}{\Delta t}
  \mbox{,}
\end{equation}
from which one can calculate the kinematic viscosity of a fluid component as 
\begin{equation}
  \label{eq:kinvis}
  \nu^c = c_S^2 \Delta t \left( \frac{\tau^c}{\Delta t} - \frac{1}{2} \right)
  \mbox{.}
\end{equation}
For convenience, the lattice and time constants are taken to be $\Delta x = \Delta t = 1$ from now on. In all simulations presented here, we have chosen $\tau^c \equiv 1$ for all components, which then implies $\nu^c = 1/6$ for all components.
The size of the simulation volume is denoted as $V_{\mathrm{box}} = n_x \cdot n_y \cdot n_z$.

\subsection{Multicomponent lattice Boltzmann}
\label{ssec:multicomponent-lb}

When further fluid species $c'$ with a single-particle distribution function $f^{c'}_i(\vec{x},t)$ are to be modeled, an interaction force $\vec{F}_{\mathrm{C}}^c(\vec{x},t)$ is calculated locally according to the approach of Shan and Chen~\cite{ bib:shan-chen:1993}:
\begin{equation}
  \label{eq:sc}
  \vec{F}_{\mathrm{C}}^c(\vec{x},t) = -\Psi^c(\vec{x},t) \sum_{c'}g_{cc'} \sum_{\vec{x}'} \Psi^{c'}(\vec{x}',t) (\vec{x}'-\vec{x})
  \mbox{ ,}
\end{equation}
with $g_{cc'}$ a coupling constant and $\Psi^c(\mathbf{x},t)$ a monotonous weight function representing an effective mass. Throughout this work, this function takes the form 
\begin{equation}
  \label{eq:psifunc}
  \Psi^c(\vec{x},t) \equiv \Psi(\rho^c(\vec{x},t) ) = 1 - e^{-\rho^c(\vec{x},t)}
  \mbox{.}
\end{equation}
This force is then incorporated into the collision term $\Omega_i^c$ in \eqnref{LBG} by adding to the velocity $\vec{u}^c(\vec{x},t)$ in the equilibrium distribution the shift 
\begin{equation}
  \label{eq:delta-u}
  \Delta \vec{u}^c(\vec{x},t) = \frac{\tau^c \vec{F}_{\mathrm{C}}^c(\vec{x},t)}{\rho^c(\vec{x},t)}
  \mbox{.}
\end{equation}
Furthermore, this forcing affects the macroscopic bulk velocity as
\begin{equation}
  \label{eq:delta-u-bulk}
  \vec{u}^c(\vec{x},t) = \frac{\sum_i f^c_i(\vec{x},t) \vec{c}_i}{\rho^c(\vec{x},t)} - \frac{1}{2} \vec{F}_{\mathrm{C}}^c(\vec{x},t)
  \mbox{.}
\end{equation}
In our case, the coupling strength $g_{cc'}$ is negative in order to obtain de-mixing and the sum over $\vec{x}'$ in \eqnref{sc} runs over all sites separated from $\vec{x}$ by one of the discrete velocities $\vec{c}_i$. In the binary fluid systems we refer to the fluid of the droplet ($d$) and the medium ($m$) as ``red'' fluid ($r$) and ``blue'' fluid ($b$), respectively. To simplify statements about the fluid:fluid ratio on lattice sites, we introduce the order parameter $\phi(\vec{x},t) = \rho^r(\vec{x},t) - \rho^b(\vec{x},t)$, referred to as ``colour''. The LB method is a diffuse interface method, with an interface width of typically $5$ lattice sites, depending weakly on the coupling strength $g_{br}$. Owing to this, there will typically also be a small but non-zero density of red fluid population in the medium and of blue fluid population inside the droplet. This will be touched upon in greater detail in section~\ref{ssec:surfacetension-theory}.

\subsection{Amphiphiles}
\label{ssec:amphiphiles}
Amphiphiles can be introduced to LB simulations in various ways. While Benzi et al. have proposed a method that can reach even vanishingly low surface tensions by including mid-range interaction forces~\cite{bib:benzi-chibbaro-succi:2009, bib:benzi-sbragaglia-succi-bernaschi-chibbaro:2009}, we avoid taking into account additional Brillouin zones and instead follow a model proposed by Chen et al.~\cite{ bib:chen-boghosian-coveney-nekovee:2000, bib:nekovee-coveney-chen-boghosian:2000, bib:furtado-skartlien:2010}. Although this method suffices to recover the qualitative behaviour of surfactant, it is limited in the surface tension reduction it can effect -- 60\% reduction being the largest achieved in our simulations. However, availability of larger reduction was deemed unnecessary for the purpose of the present work.

In addition to having its own set of distribution functions $f^s_i(\vec{x},t)$, the amphiphilic surfactant ($s$) has a dipole vector $\vec{d}(\vec{x},t)$ associated with it, representing the average orientation of the amphiphiles at a lattice site. The direction of this dipole vector can vary continuously. Its propagation is given by
\begin{multline}
  \label{eq:d-prop}
  f^s(\vec{x},t+1) \vec{d}(\vec{x},t+1) = \\ \sum_i \left( \tilde{f}^s_i(\vec{x} - \vec{c}_i,t) \tilde{\vec{d}}(\vec{x} - \vec{c}_i,t) \right)
  \mbox{.}
\end{multline}
Here, the tildes denote the post-collision values -- for a quantity $Q_i^c$: $\tilde{Q}_i^c \equiv Q_i^c + \Omega_i^c$.
The relaxation of the dipole vector can also be described by a (vector) BGK process as
\begin{equation}
  \label{eq:d-bgk}
  \tilde{\vec{d}}(\vec{x},t) = \vec{d}(\vec{x},t)  - \frac{\vec{d}(\vec{x},t) - \vec{d}^\mathrm{eq}(\vec{x},t)}{\tau^d}
  \mbox{,}
\end{equation}
with $\tau^d$ the relaxation time of the dipole orientation towards a local equilibrium $\vec{d}^{\mathrm{eq}}(\vec{x},t)$. Furthermore, the force terms as described in \eqnref{sc} are extended to account for the forces the amphiphiles exert on the red and blue fluids:
\begin{equation}
  \label{eq:sc-surf}
  \vec{F}^c(\vec{x},t) = \vec{F}_{\mathrm{C}}^c(\vec{x},t) + \vec{F}_{\mathrm{S}}^c(\vec{x},t)
  \mbox{ ,}
\end{equation}
where the lower indices denote the source of the force and $\mathrm{C}$ and $\mathrm{S}$ refer to ``colour'' and ``surfactant'', respectively. The new addition to the force term takes the form
\begin{equation}
  \label{eq:F-surf-on-fluid}
  \vec{F}_{\mathrm{S}}^c(\vec{x},t) = -2 \Psi^c(\vec{x},t)g_{cs} \sum_{i \ne 0} \tilde{\vec{d}}(\vec{x} + \vec{c}_i,t) \cdot \tensor{\boldsymbol\theta}_i \Psi^s(\vec{x} + \vec{c}_i,t)
  \mbox{ ,}
\end{equation}
where $g_{cs}$ is the force coupling constant between an ordinary and the amphiphilic species and $\tensor{\boldsymbol\theta}_i$ is a second-rank tensor defined as
\begin{equation}
  \label{eq:surf-theta}
  \tensor{\boldsymbol\theta}_i \equiv \tensor{1} - 3 \frac{\vec{c}_i \vec{c}_i}{c^2}
  \mbox{ ,}
\end{equation}
with $\tensor{1}$ the second-rank identity tensor. Similarly, the forces acting on the amphiphiles can be split into contributions from amphiphiles and ordinary fluid:
\begin{equation}
  \label{eq:F-on-surf}
  \vec{F}^s(\vec{x},t) = \vec{F}_{\mathrm{C}}^s(\vec{x},t) + \vec{F}_{\mathrm{S}}^s(\vec{x},t)
  \mbox{.}
\end{equation}
These take the forms
\begin{equation}
  \label{eq:F-fluid-on-surf}
  \vec{F}_{\mathrm{C}}^s(\vec{x},t) = 2 \Psi^s(\vec{x},t) \tilde{\vec{d}}(\vec{x},t) \cdot \sum_{c} g_{cs}\sum_{i \ne 0} \vec{\boldsymbol\theta}_i \Psi^c(\vec{x} + \vec{c}_i,t)
\end{equation}
and
\begin{multline}
  \label{eq:F-surf-on-surf}
  \vec{F}_{\mathrm{S}}^s(\vec{x},t) = -\frac{12}{\vectornorm{\vec{c}_i}^2} g_{ss} \Psi^s(\vec{x},t) \cdot \\ \sum_{i} \Psi^s(\vec{x}+\vec{c}_i,t) \bigg( \tilde{\vec{d}}(\vec{x} + \vec{c}_i,t) \cdot \vec{\boldsymbol\theta}_i \cdot \tilde{\vec{d}}(\vec{x},t) \vec{c}_i \\ + \left[ \tilde{\vec{d}}(\vec{x} + \vec{c}_i,t) \tilde{\vec{d}}(\vec{x},t) + \tilde{\vec{d}}(\vec{x},t) \tilde{\vec{d}}(\vec{x} + \vec{c}_i,t) \right] \cdot \vec{c}_i \bigg)
  \mbox{,}
\end{multline}
respectively. The coupling constant $g_{ss}$ should be negative to model attraction between two amphiphile tails and repulsion between a head and a tail. For a full derivation of these equations, cf.~\cite{ bib:chen-boghosian-coveney-nekovee:2000}.

\subsection{Nanoparticles}
\label{ssec:nanoparticles}

Nanoparticles are discretized on the lattice and coupled to both fluid species by means of a modified bounce-back boundary condition as pioneered by Ladd~\cite{ bib:aidun-lu-ding:1998, bib:ladd:1994:1, bib:ladd:1994:2, bib:ladd-verberg:2001}, resulting in a modified lattice Boltzmann equation
\begin{equation}
  \label{eq:mbb}
  f_i^c(\vec{x}+\vec{c}_i,t+1) = f^c_{\bar{i}}(\vec{x}+\vec{c}_i,t) + \Omega_{\bar{i}}^c(\vec{x}+\vec{c}_i,t) + C
  \mbox{ ,}
\end{equation}
where $C$ is a linear function of the local velocity of the particle surface, and $\bar{i}$ are defined such that $\vec{c}_i = -\vec{c}_{\bar{i}}$. Wherever $\mathbf{x}$ is occupied by a particle, \eqnref{LBG} is replaced by \eqnref{mbb}. The particle configuration is evolved in time, solving Newton's equations in the spirit of classical molecular dynamics simulations.
As the total momentum has to be conserved, an additional force acting on the particle is needed to compensate for the momentum change of the fluid caused by \eqnref{mbb}:
\begin{equation}
  \label{eq:mbb2}
  \vec{F}(t) = \big( 2f_{\bar{i}}^c(\vec{x}+\vec{c}_i,t) + C \big) \vec{c}_{\bar{i}}
  \mbox{.}
\end{equation}
As the simulation evolves in time and a particle moves around, the configuration of lattice sites occupied by the particle changes. When a site is newly occupied by a particle, the fluids on that site are deleted and their momentum is transferred to the particle through a force
\begin{equation}
  \label{eq:mom-transfer-to-ptcl}
  \vec{F}(t) = - \sum_c \rho^c(\vec{x},t) \vec{u}^c(\vec{x},t)
  \mbox{.}
\end{equation}
Lattice sites which have been newly vacated by a particle also have to be treated. In Ladd's original algorithm for a single fluid, the initial fluid density $\rho^c_{\mathrm{init}}$ would be used. However, in the case of a multicomponent system this would cause artefacts, in particular for the case of particles adsorped to an interface: fluid $b$ would be initialized where only fluid $r$ ought to be present and vice versa. To prevent such problems from occuring, a density
\begin{equation}
  \label{eq:rho-surr}
  \overline{\rho}^c(\vec{x},t) \equiv \frac{1}{N_{\mathrm{FN}}} \sum_{i_\mathrm{FN}}  \rho^c(\vec{x}+\vec{c}_{i_{\mathrm{FN}}},t)
  \mbox{,}
\end{equation}
is defined, averaged over the $N_{\mathrm{FN}}$ neighbouring fluid lattice nodes $\vec{x}_{i_{\mathrm{FN}}}$, separated from $\vec{x}$ by the velocity vector $\vec{c}_{i_{\mathrm{FN}}}$. The fluid on the vacated site is initialized with populations
\begin{equation}
  \label{eq:vacated-site}
  f^c_i(\vec{x},t) = \rho^c_{\mathrm{new}}(\vec{x},t) \cdot f^{\mathrm{eq}}_i(\vec{u}_{\mathrm{surface}}(\vec{x},t), \rho_{\mathrm{new}}(\vec{x},t))
  \mbox{,}
\end{equation}
where $\vec{u}_{\mathrm{surface}}(\vec{x},t)$ is the local velocity of the particle surface. Due to non-zero repulsive Shan-Chen forces between the particle surface and the surrounding fluid, the effective fluid density close to the particle surface might be slightly smaller than the bulk density leading to a mass drift over time if one chooses $\rho^c_{\mathrm{new}}(\vec{x},t) = \overline{\rho}^c(\vec{x},t)$. To suppress this effect we apply a correction which keeps the total mass constant on long time scales, with small fluctuations (of the order of $10^{-4}$ of the total mass) on shorter time scales~\cite{bib:jansen-harting:2011}:
\begin{equation}
  \label{eq:rho-new}
  \rho^c_{\mathrm{new}}(\vec{x},t) = \overline{\rho}^c(\vec{x},t) \left( 1 - C_0 \frac{ \sum_c \rho^c_{\mathrm{init}} }{\rho^c_{\mathrm{init}}} \frac{\Delta \rho^c(t)}{V_{\mathrm{box}}} \right)
  \mbox{,}
\end{equation}
where $\Delta \rho^c(t)$ is the total mass error of color $c$ at time $t$, and $C_0$ can be used to tune the strength of the corrections. In this work, $C_0 = 2500$ is used. To prevent instabilities, we restrict this density to be not larger or smaller than the highest and lowest surrounding density, respectively.

The potential between the particles is a Hertz potential which approximates a hard core potential and has the following form for two spheres with identical radii $r_p$~\cite{ bib:hertz:1881}:
\begin{equation}
  \label{eq:hertz-potential}
  \phi_H=K_H(2r_p-r_{ij})^{\frac{5}{2}}\quad\mbox{for}\quad r_{ij}\le 2r_p
  \mbox{,}
\end{equation}
and zero otherwise. Here, $r_{ij} \equiv \vectornorm{\vec{r}_{ij}} \equiv \vectornorm{ \vec{r}_i - \vec{r}_j}$ is the distance between the centres for two spheres $i,j$ located at $\vec{r}_i$ and $\vec{r}_j$, respectively, and $K_H$ is the force constant, which we choose to be $K_H = 100$.
Apart from the direct interaction described by the Hertz potential we correct for the limited description of hydrodynamics when two particles come very close by means of a lubrication correction. If the number of lattice points between two particles is sufficient -- at least one fluid site -- the LB algorithm reproduces the correct lubrication force automatically. If particles approach beyond this limit, the flow is no longer sufficiently resolved. The error can be corrected by an additional force term
\begin{equation}
  \label{eq:lubrication}
  \vec{F}^{\mathrm{lub}}_{ij} = - \frac{3 \pi \nu^c r_p^2 }{2} \uvec{r}_{ij} \uvec{r}_{ij} \cdot \left( \vec{u}_i - \vec{u}_j \right) \left( \frac{1}{r_{ij} - 2 r_p} - \frac{1}{\Delta_c} \right)
  \mbox{,}
\end{equation}
with $\vec{u}_i$ and $\vec{u}_j$ the velocities of particles $i$ and $j$, respectively and $\uvec{r}_{ij}$ the unit vector pointing from the centre of particle $i$ to the centre of particle $j$~\cite{ bib:ladd-verberg:2001}. Furthermore, we choose a cut-off of this lubrication force $\Delta_c = 2/3$.

The force in \eqnref{sc} also includes interactions between a lattice node outside of a particle and a lattice node inside a particle. To calculate these interactions the lattice nodes in the outer shell of the particle are filled with a ``virtual'' fluid corresponding to the density defined in \eqnref{rho-surr}: $\rho_{\mathrm{virt}}^c(\vec{x},t) = \overline{\rho}^c(\vec{x},t)$. This density is assigned to the population density $f_{\mathrm{rest}}^c(\vec{x},t)$ for which $\vec{c}_{\mathrm{rest}} = \vec{0}$. Advection and collision are not applied to this virtual fluid.

A system of two immiscible fluids and particles is considered. We define a parameter $\Delta\rho$, the particle colour, which allows to control the interaction between the particle surface and the two fluids. If $\Delta\rho$ has a positive value, we add it to the red fluid component as
\begin{equation}
  \label{eq:red-colour}
  \rho_{\mathrm{virt}}^r = \overline{\rho}^r + \Delta\rho
  \mbox{.}
\end{equation}
Otherwise we add its absolute value to the blue fluid as
\begin{equation}
  \label{eq:blue-colour}
  \rho_{\mathrm{virt}}^b = \overline{\rho}^b - \Delta\rho
  \mbox{.}
\end{equation}
By changing $\Delta\rho$ it is possible to control the contact angle $\theta_p$ of the particle. The dependence of the contact angle on the particle colour can be fitted by a linear relation, where the slope depends on the actual simulation parameters. A particle colour $\Delta\rho = 0$ corresponds to a contact angle of $\theta_p = 90^{\circ}$, i.e. a neutrally wetting particle. For a more detailed description of our simulation algorithm the reader is referred to~\cite{ bib:jansen-harting:2011}.

An alternative method to introduce solid particles to free-energy-based multicomponent LB simulations was introduced by Stratford et al.~\cite{bib:stratford-adhikari-pagonabarraga-desplat-cates:2005, bib:stratford-adhikari-pagonabarraga-desplat:2005}, while Joshi and Sun published applications of the multiphase Shan-Chen model with suspended particles~\cite{bib:joshi-sun:2009}.

\subsection{Boundary conditions}
\label{ssec:sim-method-bc}
The simulation volume is bounded at the $x = 1$ and $x = n_x$ planes by Lees-Edwards shear boundary conditions~\cite{ bib:lees-edwards:1972}, which avoid spatial inhomogeneities that occur when shear is induced by moving walls. These boundary conditions have been adapted for use in LB simulations by Wagner and Pagonabarraga~\cite{ bib:wagner-pagonabarraga:2002}, and the reader is referred to this publication for technical details. In our simulations the boundary conditions are set up in such a way as to effect a shear rate $\dot{\gamma} = 2 u_s / \left( n_x - 1 \right)$ in the $z$-direction. The remaining sides of the system are subject to ordinary periodic boundary conditions~\cite{ bib:harting-harvey-chin-venturoli-coveney:2005, bib:giupponi-harting-coveney:2006}.

\section{Surface tension}
\label{sec:surfacetension}

\subsection{Theory}
\label{ssec:surfacetension-theory}
The Young-Laplace equation relates the pressure difference $\Delta P$ over the interface between two fluids to the surface tension $\sigma$: $\Delta P = - \sigma \nabla \cdot \hat{\vec{n}}$, with $\hat{\vec{n}}$ the surface normal. For a spherical (undisturbed) droplet of one fluid of radius $R_d$ inside another fluid this equation takes the form
\begin{equation}
  \label{eq:laplace}
  \sigma = \frac{R_d \Delta P}{2}
  \hbox{.}
\end{equation}

Calculating the correct pressure jump $\Delta P = P_d - P_m > 0$ over the interface, where $P_d$ is the pressure inside the droplet and $P_m$ is the pressure in the medium, requires some care. For a single-component and single-phase system, local pressure in LB can be calculated using the simple relation $P(\vec{x}) = c_S^2 \rho(\vec{x})$ (here and in all future equations, the time dependence has been suppressed in our notation). However, when using the multicomponent Shan-Chen model for a ternary system -- consisting of simple fluid species red $r$ and blue $b$ and a surfactant species $s$ -- there is a non-zero presence of the local minority fluid throughout the system and we have to use the more complicated expression
\begin{multline}
  \label{eq:pressure}
  \frac{P(\vec{x})}{c_S^2} = \rho^{r}(\vec{x}) + \rho^{b}(\vec{x}) + \rho^{s}(\vec{x}) + \\ + \sum_{c \neq c'} g_{cc'} \Psi^{c}(\vec{x}) \Psi^{c'}(\vec{x}) + g_{ss} \Psi^s(\vec{x}) \Psi^s(\vec{x})
  \mbox{,}
\end{multline}
which takes into account pressure contributions of the fluid-fluid interactions. 

Because of the diffuse interface in LB simulations one has to make sure that the measurements are performed far enough away from the interface, so that the density is (almost) constant in the neighbourhood. We have verified that the density profiles of the system in equilibrium are flat on the inside and outside of the droplet as little as five lattice sites away from the isosurface where the colour field is zero. Hence, this effect does not cause a problem in these cases. We therefore take a spatial average of the pressure in the centre of the droplet over a small neighbourhood ($5^3$ cube of lattice sites) as $P_d$, and the local spatial average in a corner of the system (which due to the periodic boundary conditions is the furthest away one can get) as $P_m$. Densities of the fluids $c$ -- denoted as $\rho_d^c$ and $\rho_m^c$ -- can now be defined in a similar manner.

\begin{figure}
\includegraphics[width=\figwidth]{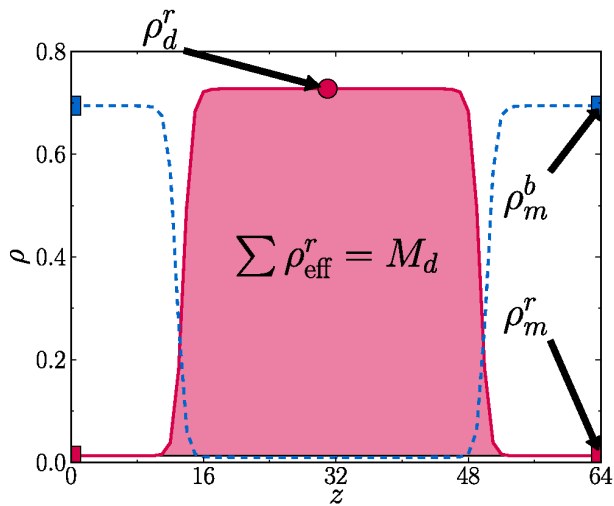}
\caption{A 1D profile of the local densities $\rho^r(\vec{x})$ (solid curve) and $\rho^b(\vec{x})$ (dashed curve) along the $z$-axis and centered in the $x-y$ plane, as used in the calculation of the droplet mass $M_d$. The droplet density of the red fluid $\rho_d^r$ and medium densities $\rho_m^c$ are taken at the center (circle) and edge (squares) of the domain, respectively. The shaded area denotes the summed effective total density of red fluid, which takes into account the non-zero ``background'' density of the red fluid in the medium $\rho_m^r$ by subtracting it from the local densities.}
\label{fig:droplet-mass-calculation}
\end{figure}

Calculating the radius of the droplet is also non-trivial, again due to the diffuse interface. We have investigated three distinct approaches, whose results have been in agreement up to less than a lattice site -- two methods based on detection of the $\phi = 0$ isosurface and one based on total mass and density of the red fluid. The latter method has been chosen, since it can most easily be extended to the case of added particles, which will be explained below. We consider the idea that all the surplus population of the red fluid ought to be contained in a sphere of constant density. We define a local effective density $\rho_{\mathrm{eff}}^r(\vec{x}) = \rho^r(\vec{x}) - \rho_m^r$ to account for the non-zero density of red fluid outside of the droplet. This effective density is used to calculate the total droplet mass
\begin{equation}
  \label{eq:droplet-mass}
  M_d = \sum_{\vec{x}} \rho_{\mathrm{eff}}^r(\vec{x})
  \hbox{.}
\end{equation}
See \figref{droplet-mass-calculation} for an illustration of this process.

Using the relation for the droplet volume $V_d = M_d / (\rho_d^r - \rho_m^r)$ and assuming sphericity of the droplet leads to
\begin{equation}
  \label{eq:droplet-radius}
  R_{d,\mathrm{mass}} = \left[ \left(\frac{3}{4 \pi}\right) \frac{M_d}{\rho_d^r - \rho_m^r} \right]^{\frac{1}{3}}
  \hbox{.}
\end{equation}

When nanoparticles are adsorped at the droplet interface (which could change the shape of the $\phi = 0$ isosurface dramatically depending on the number of particles and their position, validating the choice of this particular method), a correction term is needed to account for these particles. Recalling that the radius of the spherical particles is denoted $r_p$, we define a new effective volume of the droplet $V_d^{\mathrm{eff}} = V_d + V_p$, and approximate $V_p \approx \frac{n_p}{2} \left(\frac{4 \pi}{3} r_p^3 \right)$, where $n_p$ is the number of particles, expressing that we expect half of the particle volume to be on the inside of the interface of the droplet, adding its volume to the volume derived from the red fluid. Thus, the final equation for the radius of the droplet is given by

\begin{equation}
  \label{eq:droplet-radius-np}
  R_d = \left[ \left(\frac{3}{4 \pi}\right) \frac{M_d}{\rho_d^r - \rho_m^r} + \frac{n_p}{2} r_p^3 \right]^{\frac{1}{3}}
  \hbox{.}
\end{equation}

From \eqnref{pressure} and \eqnref{laplace} one can see that the measured surface tension depends on the fluid densities -- linearly in first order, but in a more complicated fashion in the cross terms, where the form of the effective mass function $\Psi$ plays a role (cf. \eqnref{psifunc}). In light of this, we keep the initial density of the simple fluid species constant across simulations.

\subsection{Effect of amphiphiles}
\label{ssec:surfacetension-amphiphiles}

\begin{figure}
\includegraphics[width=\figwidth]{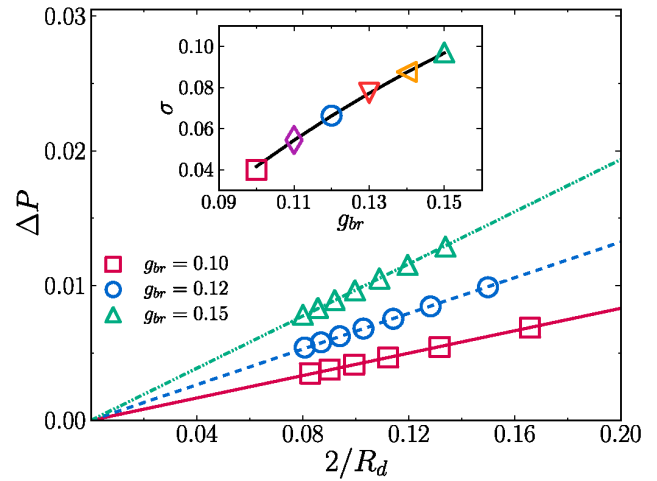}
\caption{Main plot: plotting the pressure jump over the fluid-fluid interface against the inverse droplet radius, the Young-Laplace equation for a spherical droplet allows to calculate surface tensions as the slope of fitted linear functions going through the origin $\Delta P = \sigma \left( 2/R_d \right)$ for various values of the Shan-Chen interaction parameter $g_{br}$. Inset: the surface tension $\sigma$ is a monotonically increasing function of $g_{br}$. The lines connect the results taken from the linear fits, while the symbols show the averaged results of direct evaluation of the surface tension for single points on the curves. Direct calculation of the surface tension is an accurate and efficient method when considering many systems with different parameters, which would require extra simulations with multiple droplet radii otherwise.}
\label{fig:plot-surfacetension-laplace}
\end{figure}

We now proceed to study the effect of added surfactant on the system. The system is initialized as follows: a cubic simulation volume $n_x = n_y = n_z = 64$ is considered, and the initial droplet is chosen to have a radius of $R_d^{\mathrm{init}} = 0.3 n_x = 19.2 $ and is placed in the centre of the system. These values were chosen after determining the effect of the resolution of the lattice on the surface tension. The total variation when increasing the system size from $48^3$ to $128^3$ was seen to be less than 4\%, and the best balance between accuracy and computational effort was attained at $64^3$. This error is smaller than the errors from the sources described above. After discretization, the interior droplet sites are set to have a density $\rho^r = \rho^r_{\mathrm{init}}$ and $\rho^b = 0$. Conversely, the medium sites have $\rho^r = 0$ and $\rho^b = \rho^b_{\mathrm{init}}$, while the interface is crudely modeled by a linear density gradient over 5 lattice sites. Because of stability reasons and the shape of the effective mass function $\Psi^c$ we use $\rho^r_{\mathrm{init}} = \rho^b_{\mathrm{init}} = 0.7$ in all results presented here. In the case of added surfactant, the density is set to $\rho^s = \rho^s_{\mathrm{init}}$ everywhere. The initial surfactant density varies from simulation to simulation and will always be reported explicitly. As the system approaches its equilibrium state, surfactant accumulates at the interface, causing the local density at the interface to be higher by a factor of approximately two compared to the density in the bulk. Reaching the equilibrium state from this initialization can take a long time -- to obtain stable results for the surface tension the simulations have to run for tens of thousands of time steps for the systems described in this paragraph (and up to several hundred thousand timesteps in the case of a system with particles, as will be described in section~\ref{ssec:surfactension-nanoparticles}).

Firstly, we are interested in determining the surface tension as a function of the fluid-fluid interaction strength $g_{br}$ in the case of a binary fluid system. We fix the coupling constants related to the surfactant to limit the parameter space of interest and choose $g_{rs} = g_{bs} = g_{ss} = -0.005$. As discussed in section~\ref{ssec:amphiphiles}, these have to be negative to properly model the behaviour of a surfactant. The actual values are chosen for their stability. There are also some restrictions on our choice of $g_{br}$. For $g_{br} < 0.10$ the fluids become miscible when surfactant with the properties specified above is added, leading to ill-defined interfaces and droplets. Furthermore, choosing $g_{br} > 0.15$ leads to numerical instabilities~\cite{ bib:schmieschek-harting:2011}. We therefore consider the values $0.10 \le g_{br} \le 0.15$, restricting reachable surface tensions. Rewriting Eqn.~\ref{eq:laplace} as $\Delta P = \sigma \left( 2 /R_d \right)$ allows to extract $\sigma$ by considering it to be the slope of the pressure difference plotted against twice the inverse droplet radius. Linear fits through the origin correspond very well to the simulation results for $0.10 \le g_{br} \le 0.15$ (cf. \figref{plot-surfacetension-laplace}). From this it follows that $g_{br}$ can be mapped onto the surface tension: $\sigma \equiv \sigma(g_{br})$, with $\sigma(g_{br})$ a monotonically increasing function. The inset of~\figref{plot-surfacetension-laplace} shows that calculating surface tensions directly using a single droplet radius together with Eqn.~\ref{eq:laplace} is an accurate and efficient method that does not require multiple simulations for a single choice of $g_{br}$.

\begin{figure}
\includegraphics[width=\figwidth]{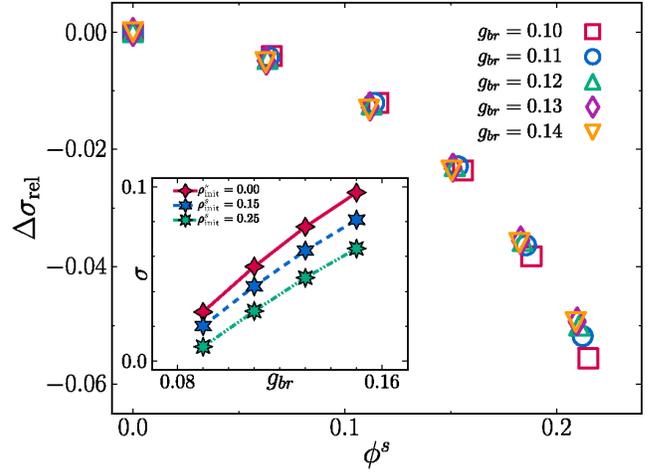}
\caption{Main plot: by rescaling the effect of surfactant to $\Delta \sigma_{\mathrm{rel}} \equiv \left( \sigma / \sigma_0 - 1 \right) g_{br}$ the curves for different values of fluid-fluid interaction strength $g_{br}$ can be made to collapse, illustrating the fact that the effect of added surfactant scales with $g_{br}$. The error bars of the data points are too small to be visible at this scale. Inset: surface tensions as a function of $g_{br}$ for various concentrations of surfactant $\rho^s_{\mathrm{init}}$. The lines are not a fit, but included only to guide the eye. This shows qualitatively that the surfactant lowers the surface tension, as expected. Again, the error bars are too small to be visible in this plot.}
\label{fig:plot-surf-sigma-eq-g_br}
\end{figure}

The qualitative result of creating a ternary system by adding an amphiphilic surfactant component to the binary droplet system is as expected: increasing surfactant density from $\rho^s_{\mathrm{init}} = 0.0$ to $\rho^s_{\mathrm{init}} = 0.15$ and $\rho^s_{\mathrm{init}} = 0.25$ lowers the surface tension by $30$ to $50$ percent (cf. the inset of \figref{plot-surf-sigma-eq-g_br}). As mentioned in section~\ref{ssec:amphiphiles}, this relatively modest reduction is due to limitations of the surfactant model used for these simulations. It is, however, sufficient to highlight the differences between the effect of amphiphiles and nanoparticles. To find a quantitative relation between surfactant concentration fraction $\phi^s \equiv \rho_{\mathrm{init}}^s / \left( \rho_{\mathrm{init}}^s + \rho_{\mathrm{init}}^b \right)$, interaction strength and surface tension, it is useful to define
\begin{equation}
  \label{eq:delta-sigma-rel}
  \Delta \sigma_{\mathrm{rel}} \equiv \left( \frac{\sigma}{\sigma_0} - 1 \right) g_{br}
  \hbox{,}
\end{equation}
where $\sigma_0 \equiv \sigma(\phi^s = 0)$.
Plotting this quantity as a function of $\phi^s$, the data points collapse onto a universal curve, as shown in \figref{plot-surf-sigma-eq-g_br}. This illustrates the fact that the effect of the surfactant scales with the interaction strength between the two non-amphiphilic fluid species. We can use this data to obtain another mapping: $\sigma \equiv \sigma(g_{br},\phi^s)$ for fixed interaction strengths involving the surfactant species. These mappings will later be used in determining capillary numbers for systems of a droplet subjected to shear.

\subsection{Effect of nanoparticles}
\label{ssec:surfactension-nanoparticles}

\begin{figure}
\includegraphics[width=\figwidth]{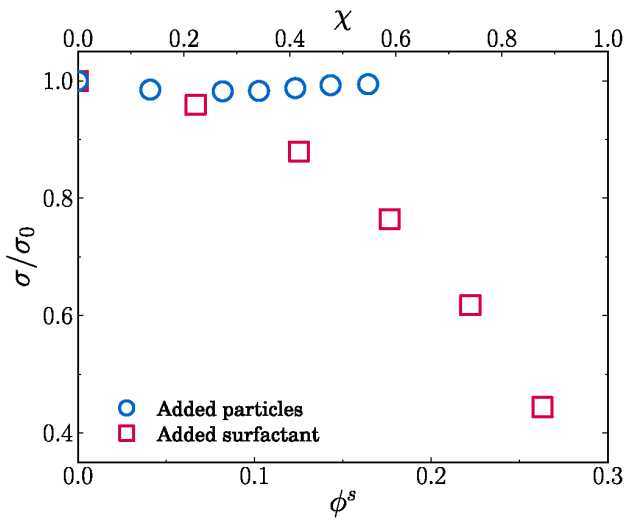}
\caption{Surface tension change as a function of particle droplet surface coverage $\chi$ (top $x$-axis, circles) and surfactant volume fraction $\phi^s$ (bottom $x$-axis, squares). Here, $\sigma_0$ is the surface tension for the purely binary system (i.e. $\chi = 0$ and $\phi^s = 0$, respectively) with otherwise identical parameters. For all cases $g_{br} = 0.10$; for the system with surfactant $g_{rs} = g_{bs} = g_{ss} = -0.005$ and for the system with particles $r_p = 5.0$, $m_p = 524$, and $\theta_p = 90^\circ$. Introducing 25\% volume fraction of surfactant into the system lowers the surface tension by almost 60\%, while particles affect it only very weakly. The slight drop in measured surface tension for moderate values of $\chi$ is caused by errors introduced in the calculation of the droplet radius due to anisotropic particle distributions on account of spurious currents at the droplet interface.}
\label{fig:plot-surfacetension-fg_chi-rescaled}
\end{figure}

Next, the case of added (spherical and monodisperse) nanoparticles is considered. The fraction $\chi$ of the droplet surface removed by the adsorped particles is a parameter of interest. The excluded surface area due to one neutrally wetting particle is a spherical cap whose area is given by $A_p^{\mathrm{ex}} = 2 \pi R_d \left( R_d - \sqrt {R_d^2 - r_p^2} \right)$, from which follows that the total coverage fraction of a spherical droplet is given by
\begin{equation}
  \label{eq:chi}
  \chi \equiv n_p \frac{A_p^{\mathrm{ex}}}{A_d} = n_p \frac{R_d - \sqrt {R_d^2 - r_p^2}}{2R_d}
  \hbox{.}
\end{equation}
Since we use a diffuse interface method, any suspended particles have to be of sufficient size compared to the interface width to resolve their interfacial properties. In practice, this means a typical spherical particle needs to have a diameter of at least $10$ LB length units, while a spherical droplet should be larger than the particles by an order of magnitude. Allowing then sufficient room for the deformation of the droplets to take place without undue finite size effects, these calculations remain computationally challenging, even for the case of a single droplet and a highly efficient massively parallel simulation environment. In order to be able to let the droplet deform sufficiently in later simulations we also elongate the system in the direction of the shear flow ($z$-direction): $n_x = n_y = 256$, $n_z = 512$. The droplet is initialized as described above, with initial radius $R_d^{\mathrm{init}} = 0.3 \cdot n_x = 76.8$ and we choose $g_{br} = 0.10$. The particles have a radius $r_p = 5.0$ and are neutrally wetting ($\theta_p = 90^{\circ}$). Furthermore, they have a mass $m_p = 524$, which corresponds to a density $\rho^p = 1$ (taken with respect to the lattice). They are initialized on a spiral running over the surface of the initial droplet from the north to south pole, resulting in a very uniform initial distribution of particles at low computational cost~\cite{ bib:bauer:2000}. When the system is allowed to get into its equilibrium state, however, some pattern formation of the particles on the interface occurs, due to the occurence of spurious currents near the interface (as also observed in similar modeling of liquid-vapour systems by Joshi and Sun~\cite{ bib:joshi-sun:2009}). This effect is negligible when the system is not stationary: the currents are much smaller than the effect of applying shear to the system, or, for example, the effect of droplet movement in the formation of a Pickering emulsion. In either case the particle ordering due to the spurious currents is destroyed. 

Adding particles with the aforementioned properties does not affect surface tension at all -- the presence of particles at the interface only changes interfacial free energy directly by taking away energetically expensive fluid-fluid interface and replacing it with cheap particle-fluid interfaces. To clarify this, consider the free energy term $F_{\sigma}$ related to the surface tension of the interface of the droplet $D$,
\begin{equation}
  \label{eq:free-energy}
  F_{\sigma} = \oint\limits_{\partial D} \sigma \, \mathrm{d} A
  \hbox{,}
\end{equation}
which integrates the surface tension over the interface. For simplicity, the surface tension is taken to be constant over the interface. There are now two possibilities to reduce this energy. The first is to reduce the surface tension $\sigma$, which is the effect of added surfactant. Because $\sigma > 0$ and the integration only pertains to the fluid-fluid interface, the second possibility is to reduce the area of integration $\partial D$, which is effected by adsorped particles. The particles also add energy to the system by means of the interfacial energy between the particle and either fluid, but as this energy per unit surface area is much smaller than the fluid-fluid surface tension, the net effect is still a reduction of the free energy.

A comparison of the addition of surfactant and nanoparticles to a binary system can be seen in \figref{plot-surfacetension-fg_chi-rescaled}. Due to the anisotropic distribution of the particles on the interface, errors are introduced in the calculation of the droplet radius for intermediate values of $\chi$, lowering the measured surface tension by up to $3\%$. At higher $\chi$, the anisotropy disappears, and with it the calculated change in surface tension, which returns to its original value for $\chi \approx 0.5$.  In the system with surfactant an identical value of $g_{br} = 0.10$ is used. Unlike adding particles, adding surfactant lowers the surface tension (a 60\% drop in surface tension for $\rho^s_{\mathrm{init}} = 0.25$).

\section{Droplet in shear flow}
\label{sec:deformation}

\subsection{Theory}
\label{ssec:deformation-theory}

\begin{figure}
\includegraphics[angle=0,width=\figwidth]{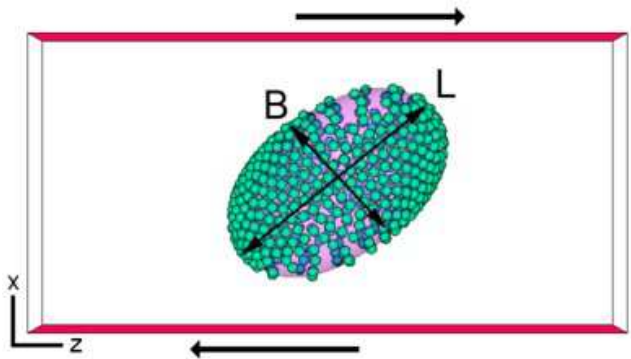}
\caption{Representative deformation of a particle-covered droplet at $\mathrm{Ca} = 0.075$, $\chi = 0.55$. The shaded planes at the top and bottom are subject to Lees-Edwards boundary conditions, inducing a shear rate $\dot{\gamma} = 2 u_s / \left( n_x - 1 \right)$ in the $z$-direction, as discussed in section~\ref{ssec:sim-method-bc}. The shear causes droplet deformation $D \equiv (L -B) / (L + B)$ and an inclination of the droplet of angle $\theta_d$, which is the angle the long axis of the droplet $L$ forms with the shear direction $z$.}
\label{fig:deformed-droplet-shear}
\end{figure}

The system of a droplet of a fluid suspended in another fluid is subjected to simple shear flow, which causes the droplet to deform (cf. \figref{deformed-droplet-shear}). To analyze this process we first define a set of dimensionless variables. 
The dimensionless deformation parameter
\begin{equation}
  \label{eq:taylor}
  D \equiv \frac{L-B}{L+B}
\end{equation}
introduced by Taylor~\cite{bib:taylor:1932,bib:taylor:1934} is used to describe the deformation of the droplet, where $L$ is the length and $B$ is the breadth of the droplet. If the droplet is a perfect prolate ellipsoid the length and breadth can be related to the long and short axes, respectively, but in other cases a length and breadth of a more irregular shape can still be determined. One can easily see that for a spherical droplet $L = B$, hence $D = 0$, and for a strongly deformed droplet, $L \gg B$, $D \to 1$. Extraction of $D$ from the data is effected through the symmetric moment of inertia tensor
\begin{equation}
  \label{eq:moment-of-inertia}
  \tensor{I} = \begin{bmatrix} I_{11} & I_{12} & I_{13} \\ I_{12} & I_{22} & I_{23} \\ I_{13} & I_{23} & I_{33} \end{bmatrix}
  \hbox{.}
\end{equation}
In order to define these moments of inertia, we first calculate the centre-of-mass position of the droplet
\begin{equation}
  \label{eq:com}
  \vec{x}_d^{\mathrm{com}} = \sum_{\vec{x} \in V_{\mathrm{box}}} \vec{x} \cdot \rho_{\mathrm{com}}^r(\vec{x}) 
  \hbox{,}
\end{equation}
where a cutoff density $\rho_{\mathrm{cutoff}}^r$ is introduced to confine the summation to the droplet: 
\begin{equation}
  \label{eq:rho-hat}
  \rho_{\mathrm{com}}^r(\vec{x}) = 
  \begin{cases}
    \rho^r(\vec{x}) & \text{if $\rho^r(\vec{x}) > \rho_{\mathrm{cutoff}}^r$} \\
    0 & \text{otherwise.}
  \end{cases}
\end{equation}
The cutoff density should fulfill the condition $\rho_m^r < \rho_{\mathrm{cutoff}}^r < \rho_d^r$ and can be chosen freely within that range with negligible effect on the subsequent calculations. We use $\rho_{\mathrm{cutoff}}^r = 0.1$ in this work. A droplet mass based on the density $\rho_{\mathrm{com}}^r(\vec{x})$ is introduced for later use:
\begin{equation}
  \label{eq:M-com}
  M_d^{\mathrm{com}} = \sum_{\vec{x} \in V_{\mathrm{box}}} \rho_{\mathrm{com}}^r(\vec{x}) 
  \hbox{.}
\end{equation}
Defining $\tilde{\vec{x}} \equiv \vec{x} - \vec{x}_d^{\mathrm{com}}$ allows to express the elements of $\tensor{I}$ as
\begin{equation}
  \label{eq:Iij-def}
  I_{ij} = \sum_{\vec{x} \in V_{\mathrm{box}}} \rho_{\mathrm{com}}^r(\vec{x}) \left( \vectornorm{\tilde{\vec{x}}}^2 \delta_{ij} - \tilde{x}_i \tilde{x}_j \right)
  \hbox{,}
\end{equation}
where $\delta_{ij}$ is the Kronecker delta. The moment of inertia tensor $\tensor{I}^{\mathrm{ell}}$ of an ellipsoid of uniform density is a diagonal matrix with its non-zero elements given by
\begin{equation}
  \label{eq:I-ell}
  I_{ii}^{\mathrm{ell}} = \frac{M^{\mathrm{ell}}}{5} \left( \left( 1 - \delta_{i1} \right) a^2 + \left( 1 - \delta_{i2} \right) b^2 + \left( 1 - \delta_{i3} \right ) c^2 \right)
  \hbox{,}
\end{equation}
with $M^{\mathrm{ell}}$ the mass of the ellipsoid and $a$, $b$ and $c$ the length of the axes. We now assume that the deformed droplet can be approximated by such an ellipsoid, and $M^{\mathrm{ell}} = M_d^{\mathrm{com}}$. The set of equations obtained by combining the eigenvalues of $\tensor{I}$ with~\eqnref{I-ell} can be solved for $a$, $b$ and $c$. The length and breadth of the droplet are then defined as $L = \max(a,b,c)$ and $B = \min(a,b,c)$, respectively.

A droplet thus deformed has lost its spherical shape and gains a preferred alignment. This is expressed through the inclination angle $\theta_d$. It is calculated by taking the eigenvector $\vec{L}$ corresponding to the long axis of the droplet of the moment of intertia tensor $\tensor{I}$, and calculating the arctangent of the quotient of its $x$ and $z$ components:
\begin{equation}
  \label{eq:theta_d}
  \theta_d = \arctan{ \frac{L_x}{L_z} }
  \hbox{.}
\end{equation}

\begin{figure}
\includegraphics[width=\figwidth]{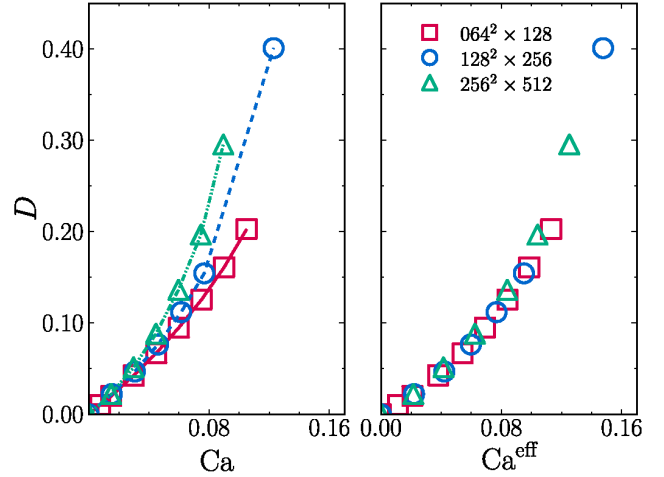}
\caption{Dimensionless deformation $D$ of a droplet in shear flow as a function of capillary number $\mathrm{Ca} \equiv \mu_m \dot{\gamma} R_d / \sigma$ (left) and effective capillary number $\mathrm{Ca}^{\mathrm{eff}} \equiv \mu_m \dot{\gamma}^{\mathrm{eff}} R_d / \sigma$ (right). Different symbols represent different system sizes. The capillary number computed from an assumed undisturbed shear profile gives rise to divergence in the relation between $\mathrm{Ca}$ and the deformation when the system size changes (lines are included to guide the eye). However, these points collapse on the curve which uses the effective capillary number, which takes into account the actual shear experienced by the droplet.}
\label{fig:plot-sheared-effective-Ca}
\end{figure}

A capillary number $\mathrm{Ca}$ can be defined as $\mathrm{Ca} \equiv \mu_m \dot{\gamma} R_d / \sigma$, where $\mu_m$ is the dynamic viscosity of the medium, $\dot{\gamma}$ is the shear rate as imposed through the Lees-Edwards boundary conditions, $R_d$ is the radius of the initial -- undeformed, hence spherical -- droplet and $\sigma$ is the surface tension. However, using this definition of the capillary number does not take into account the substantial distortion of the linear shear gradient caused by the presence of the droplet, which leads to a dependence on the size of the simulation volume, even in the case when only the resolution of the simulation is increased. 
\begin{figure}
\includegraphics[width=\figwidth]{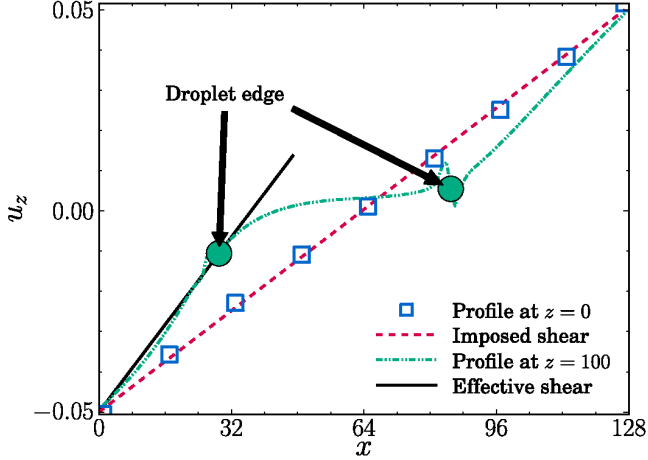}
\caption{Representative $z$-velocity profiles of a droplet with initial radius $R_d = 39.2$, centered in a system with $n_x = n_y = 128$, $n_z = 256$ and $u_s = 0.05$. The cuts are taken in $x$-direction at $y = 63$ and taken through the droplet ($z = 100$) as well as at the edge of the periodic volume ($z = 0$). Also shown is what the imposed shear rate would look like in absence of the droplet ($\dot{\gamma}$ is the slope of this line). It is clear that far away from the droplet, the measured shear is almost undisturbed and linear, while the droplet locally strongly disturbs the effective shear profile. We detect the droplet interface and calculate an effective shear rate $\dot{\gamma}^{\mathrm{eff}}$ based on the slope of the profile in the region between the shear boundary and the droplet. Because of the deformation and inclination of the droplet this curve will generally not be symmetrical for the top and bottom shear boundaries for any particular given value of $z$, however, averaging over the length of the droplet restores this symmetry.}
\label{fig:vel-profile-example}
\end{figure}
A better characterization of the system can therefore be found in an effective capillary number:
\begin{equation}
  \label{eq:Ca-eff}
  \mathrm{Ca}^{\mathrm{eff}} \equiv \frac{\mu_m \dot{\gamma}^{\mathrm{eff}} R_d}{\sigma}
  \hbox{,}
\end{equation}
where an effective shear rate $\dot{\gamma}^{\mathrm{eff}}$ is measured in the simulation, instead of assuming the validity of an imposed shear rate set directly by an input parameter. \figref{vel-profile-example} depicts the measurement of $\dot{\gamma}^{\mathrm{eff}}$ for a droplet with initial radius $R_d = 39.2$ in a system of size $n_x = n_y = 128$, $n_z = 256$ and with $u_s = 0.05$. Far away from the droplet $\dot{\gamma}^{\mathrm{eff}} \approx \dot{\gamma}$, but for those values of $z$ over which the droplet extends, typically $\dot{\gamma}^{\mathrm{eff}} > \dot{\gamma}$. The slope of the velocity gradient between the shear boundary and the droplet interface can be measured, which is then averaged over the length of the droplet to obtain the effective shear rate. This shear rate better characterizes the system. When the effective capillary number is used, taking into account the actual shear experienced by the droplet, the dependence of the deformation on the system size disappears, as shown in \figref{plot-sheared-effective-Ca}, where deformations of a droplet are plotted against both $\mathrm{Ca}$ and $\mathrm{Ca}^{\mathrm{eff}}$. When the original capillary number is used, the deformation curves diverge as the system size increases from $64^2 \cdot 128$ to $128^2 \cdot 256$ and $256^2 \cdot 512$, while the curves collapse when plotted as a function of the effective capillary number.

We also define the ratio of the droplet and medium viscosity $\lambda \equiv \mu_d / \mu_m = 1$ in all presented data, as well as a Reynolds number $\mathrm{Re} \equiv \frac{\rho_m \dot{\gamma} R_d^2}{\mu_m}$ and an effective Reynolds number
\begin{equation}
  \label{eq:Re-eff}
  \mathrm{Re}^{\mathrm{eff}} \equiv \frac{\rho_m \dot{\gamma}^{\mathrm{eff}} R_d^2}{\mu_m}
  \hbox{.}
\end{equation}
Due to the variation in system size and shear rate, the Reynolds number varies between approximately $0.6 < \mathrm{Re}^{\mathrm{eff}} < 25$, the effect of which will be discussed in section~\ref{ssec:deformation-inclination-results}.

\subsection{Distribution of amphiphiles and nanoparticles}
\label{ssec:distribution}

\begin{figure}
\includegraphics[width=\figwidth]{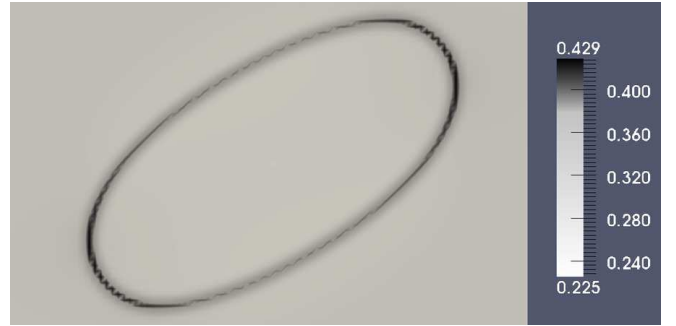}
\caption{Distribution of surfactant in a system of size $n_x = n_y = 64$, $n_z = 128$. The local surfactant density $\rho^s(\vec{x})$ is plotted on a 2D cut showing the centered $x$-$z$ plane through a droplet sheared with constant velocity $u_s = 0.06$ and $\rho^s_{\mathrm{init}} = 0.25$ ($\mathrm{Ca}^{\mathrm{eff}} = 0.16$). The snapshot is zoomed into the droplet and as such does not accurately reflect confinement of the droplet or the elongation of the system. Surfactant accumulates at the droplet interface until saturation occurs. Compared to the remainder of the interface, a slightly higher local density is observed at the tips of the droplet (10 to 20 percent). This is caused by convection of the surfactant~\cite{ bib:stone-leal:1990}. }
\label{fig:picture-surfactant-snapshot}
\end{figure}

To understand the effect of amphiphiles and nanoparticles on the deformation properties of the droplet, we first discuss how they position themselves at and move over the droplet interface as the droplet is sheared.

The distribution of surfactant on a 2D cut through a sheared droplet is shown in~\figref{picture-surfactant-snapshot}. In this example, the shear rate is held constant at $\dot{\gamma} = 0.002$ and the initial surfactant density is set to $\rho^s_{\mathrm{init}} = 0.25$. As has been mentioned in section~\ref{ssec:surfacetension-amphiphiles}, the surfactant accumulates at the interface. When the system is subjected to shear, a slightly increased density of approximately 10 to 20 percent is observed at the tips of the droplet, due to convection of the surfactant~\cite{ bib:stone-leal:1990}. This behaviour is more readily apparent for lower $\rho^s_{\mathrm{init}}$ and is different from our observations in the case of adsorped particles, as we will show below.

\begin{figure}
\begin{tabular}{l l l l }
a)\vspace*{-0.3cm} & & & \\
&\includegraphics[width=0.27\linewidth]{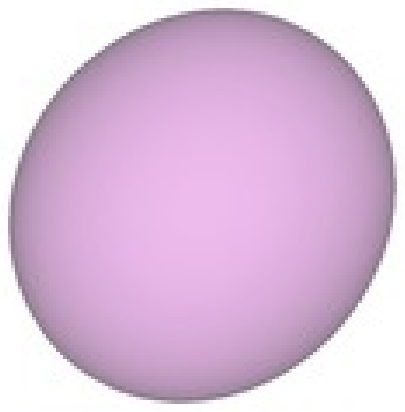} &
\includegraphics[width=0.27\linewidth]{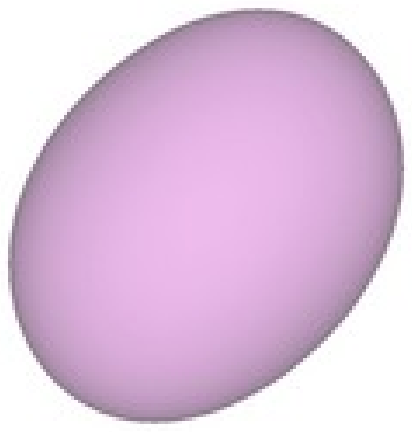} &
\includegraphics[width=0.27\linewidth]{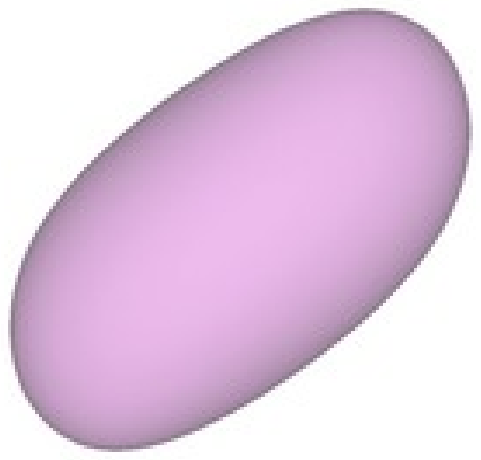} \\
b)\vspace*{-0.3cm} & & & \\
&\includegraphics[width=0.27\linewidth]{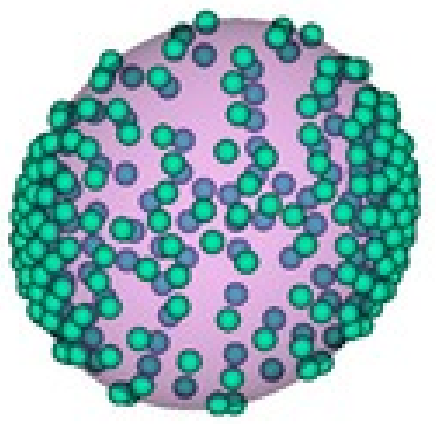} &
\includegraphics[width=0.27\linewidth]{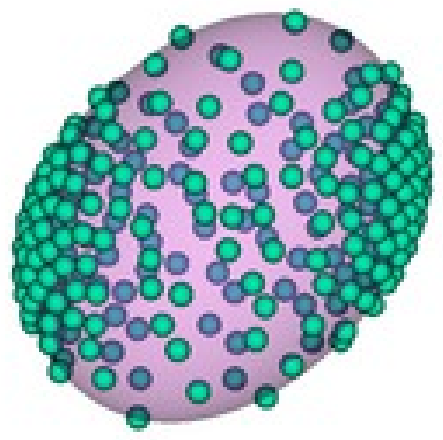} &
\hspace{-0.4cm}\includegraphics[width=0.27\linewidth]{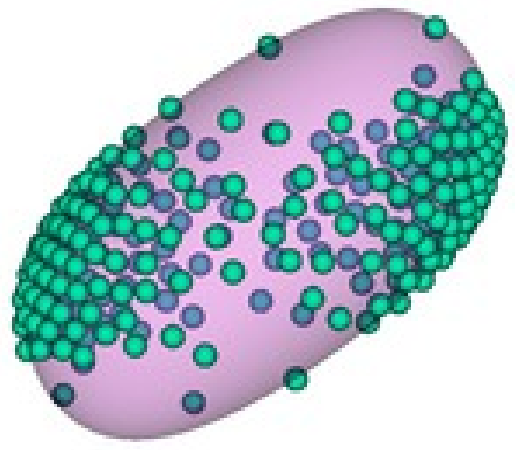} \\
c)\vspace*{-0.3cm} & & & \\
&\includegraphics[width=0.27\linewidth]{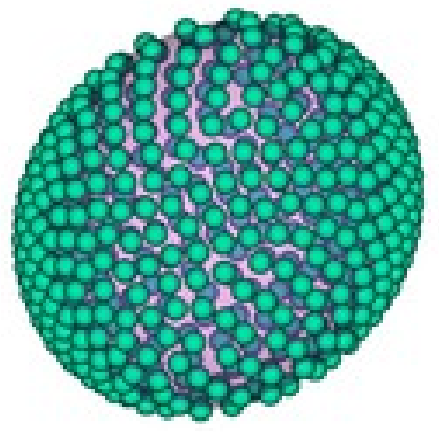} &
\includegraphics[width=0.27\linewidth]{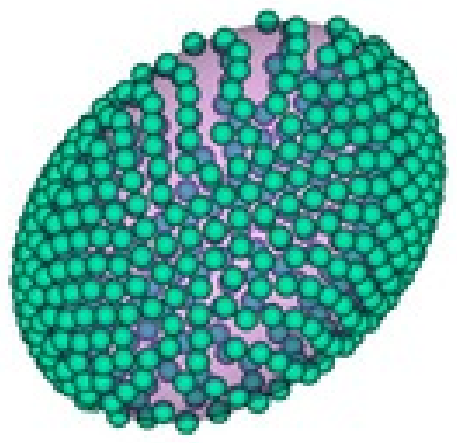} &
\includegraphics[width=0.27\linewidth]{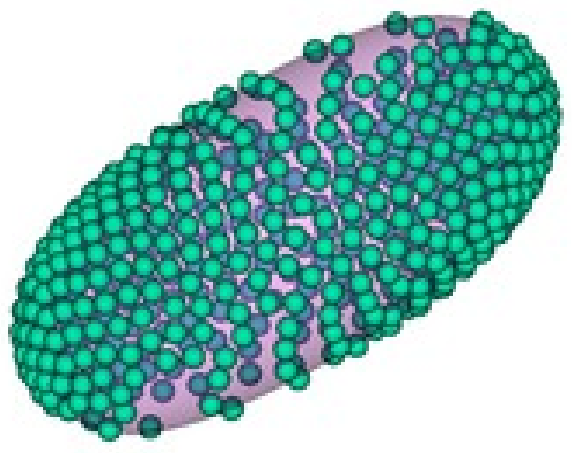} \\
& $\mathrm{Ca}^{\mathrm{eff}} = 0.04$ & $\mathrm{Ca}^{\mathrm{eff}} = 0.08$ & $\mathrm{Ca}^{\mathrm{eff}} = 0.12$ \\
\end{tabular}
\caption{Side-view examples of deformed droplets, for various particle coverage fractions: a) $\chi = 0.00$, b) $\chi = 0.27$ and c) $\chi = 0.55$. In these pictures the shear velocities are horizontal. In all these simulations $g_{br} = 0.10$, $r_p = 5.0$, $m_p = 524$, and $\theta_p = 90^\circ$. One can see that although increasing $\chi$ from $0$ to $0.27$ does not strongly change the deformation of the droplet, the particles themselves do exhibit interesting behaviour: they prefer to stay in the middle of the channel where the shear flow is weakest (recall that the top and bottom planes are moving inducing flow in opposite directions). This causes the formation of a band of particles near the equator of the droplet, with the axis through the poles in $x$-direction. For packings of higher density there is an interplay between shear forces and the curvature of the interface, which causes the aforementioned band to grow asymmetrically as the particles prefer to occupy interface with high local curvature. The particles also exhibit tank-treading-like behaviour: they move around the interface following the shear flow. The combined effect of this tank-treading-like movement and the energy arguments described above lead to the formation of strings of single particles, being swept from the band near one tip to the other tip.}
\label{fig:pictures-sheared-particles}
\end{figure}

\begin{figure}
\includegraphics[width=\figwidth]{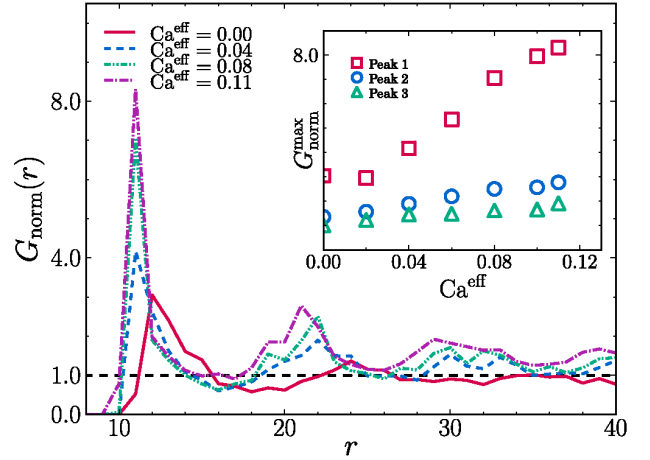}
\caption{Main plot: normalized pair correlation function between particles $G_{\mathrm{norm}}(r)$, for $\chi = 0.41$, $r_p = 5.0$ and various capillary numbers. As the capillary number increases, the peaks both shift in position and increase in height. The former effect is an indication of closer packing, while the latter corresponds to the observation of preferred regions for the particles as shown in \figref{pictures-sheared-particles}~b). Inset: the height of the first three peaks of the normalized pair correlation function are shown as a function of the effective capillary number. The strongest effect is seen for the very first peak, which shows the largest growth in both relative and absolute sense and increases in height by almost a factor of 3.}
\label{fig:plot-pair-correlations-inset}
\end{figure}

Even if the droplet interface is initially densely packed with particles, this will no longer be the case when the droplet deforms -- the interfacial area increases while the number of particles remains constant. The particles then have freedom to move over the interface to some extent (cf.~\figref{pictures-sheared-particles}). In all cases, however, detaching particles from the interface remains practically impossible. The particles are swept over the interface with increasing velocity as they move away from the centre plane of the system and up the shear gradient. If the particles would not be affected by the shear flow, they would prefer to occupy interface with high local curvature as can be explained by a geometrical argument: the interface removed by a spherical particle at a curved interface is larger than the circular area removed from a flat interface, and this effect gets stronger as curvature increases. This explains why in this dynamic equilibrium, most particles can be found at the tips of the droplet. This can be observed in \figref{pictures-sheared-particles}~b) at high capillary number, where the relatively flat sections of the interface at the top and bottom of the droplet have the lowest particle density and the strongly curved section of the interface near the centre plane is much more highly populated than the strongly curved section protruding farther into the shear flow.

To quantify these phenomena, we employ a discrete pair correlation function
\begin{equation}
  \label{eq:pair-correlation}
  G(r) = \sum_{i = 1}^{n_p} \sum_{j = 1}^{i - 1} \int_r^{r+1} \delta( \vectornorm{\vec{r}_i - \vec{r}_j} - R) \, \mathrm{d} R
  \hbox{,}
\end{equation}
where $\delta(x)$ is the Dirac delta function. Because of the system size, the domain of the pair correlation function is limited to $0 \le r < n_x / 2 = 128$. Furthermore, we choose to employ $G_{\mathrm{norm}}(r)$, which is proportional to $G(r)$ and is normalized to tend to $1$ as $r$ tends to its maximum value. In \figref{plot-pair-correlations-inset} we show this normalized pair correlation function for $\chi = 0.41$, $r_p = 5.0$ and various capillary numbers. In the main plot, two features are readily apparent: the peaks of the function both shift in position and increase in height as the capillary number is increased. The former effect indicates a denser overall packing of the particles, which occurs despite the fact that more interfacial area becomes available as the droplet deforms. The latter effect corresponds to the emergence of preferred regions for particles as described above. In the inset we show the height of the first three peaks of $G_{\mathrm{norm}}(r)$. The first peak shows the largest increase, both in absolute and relative sense. This is caused by the fact that at $\chi = 0.41$ the band of particles around the droplet is not dense everywhere, but is mostly restricted to patches near the tips of the droplet. Thus, particles having a close packing around them extending over more than one particle distance (which would show peaks of higher order) are more rare than those with just closely packed neighbours. Finally, we have observed that when this structure is established it is stable over time for as long as the system is subjected to a constant shear. When this shear is removed, the particles restore themselves to their former pattern, as described in section~\ref{ssec:surfactension-nanoparticles}, just as the droplet shape returns to that of a sphere.

\begin{figure}
\includegraphics[width=\figwidth]{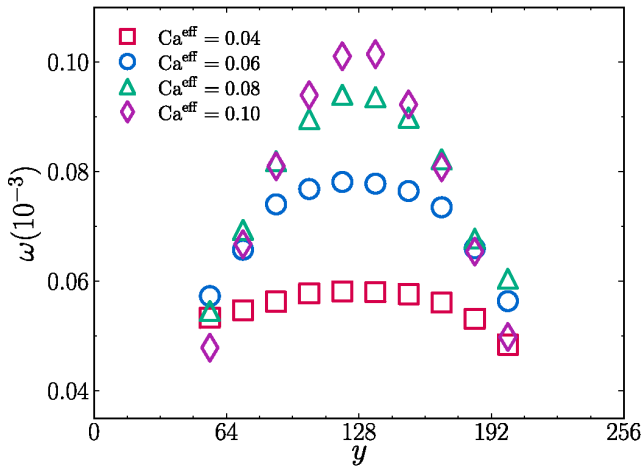}
\caption{Rotational frequency of particles $\omega$ as a function of their position on the $y$-axis for $\chi = 0.55$ and $r_p = 5.0$ and various capillary numbers. The rotational frequency does not explicitly take into account the increase in the circumference of a cut of the droplet perpendicular to the $y$-axis due to deformation. As the capillary number is increased through increased shear rate, the particles' rotation frequency increases, with particles in the middle of the droplet being both the fastest and getting the largest speedup, despite the fact that the particles in the middle travel the longest paths. The differences in frequency highlight that the movement is different from tank-treading as observed in, for example, vesicles~\cite{ bib:kaoui-harting-misbah:2011}.}
\label{fig:plot-sheared-movement-np512}
\end{figure}

Even though the overall structure of the particles on the droplet interface remains stable over time, individual particles move over the interface, performing a quasi-periodic motion. Their trajectories follow the motion of the shear flow and loop around the droplet with a rotational frequency $\omega$. We demonstrate in \figref{plot-sheared-movement-np512} that this frequency is not constant for all particles, instead showing a dependence on the position of the particle along the $y$-axis. When deformation is considered for ellipsoidal cuts of the droplet along the $y$-axis, the deformation is highest in the centre of the droplet, giving particles greater options for mobility that are also better-aligned with the shear flow, leading to increased particle velocities. This is qualitatively different from the tank-treading behaviour observed in, for example, vesicles~\cite{ bib:kaoui-harting-misbah:2011}, which is characterized by a constant frequency for all points. We also observe that the average rotation frequency increases with increasing capillary number, in spite of the fact that the particles need to follow longer paths to complete one revolution as the droplet deforms. This increase in frequency is concentrated on the particles in the centre of the droplet, for the same reasons as mentioned above.

\subsection{Droplet deformation and inclination}
\label{ssec:deformation-inclination-results}

\begin{figure}
\includegraphics[width=\figwidth]{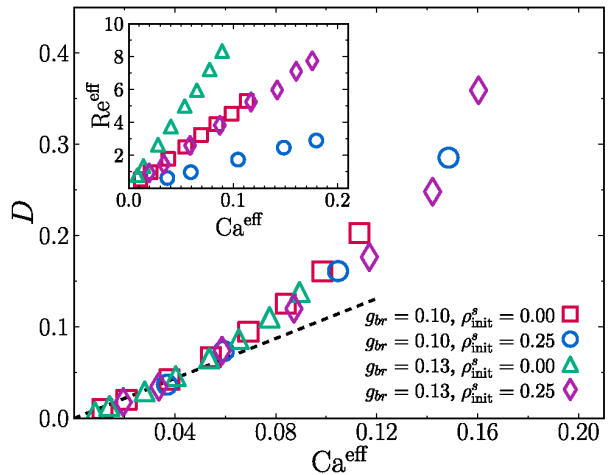}
\caption{Main plot: deformation parameter $D$ as a function of the effective capillary number $\mathrm{Ca^{\mathrm{eff}}}$ for various interaction strengths $g_{br}$ and surfactant densities $\rho^s$. The system size is $n_x = n_y = 64$, $n_z = 128$ and the surfactant interaction strenghs are fixed at $g_{rs} = g_{bs} = g_{ss} = -0.005$. The capillary number is varied by changing the shear rate. Over the entire domain, the curves collapse onto a universal curve, and for small capillary number Taylor's law is recovered (dashed line). Note that the squares in this plot correspond to the squares in \figref{plot-sheared-effective-Ca}. Inset: The relation between the capillary and Reynolds numbers is a linear one. The slope is proportional to the surface tension (cf. \eqnref{Re-Ca}).}
\label{fig:plot-sheared-surfactant-deformation}
\end{figure}

\begin{figure}
\includegraphics[width=\figwidth]{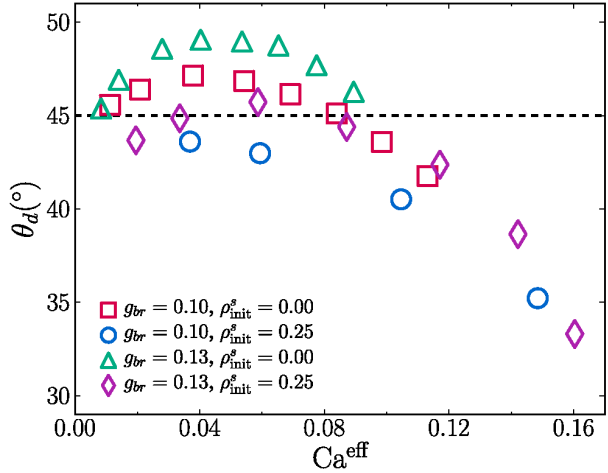}
\caption{Inclination angle $\theta_d$ of a droplet as a function of the effective capillary number $\mathrm{Ca^{\mathrm{eff}}}$ for various interaction strengths $g_{br}$ and surfactant densities $\rho^s$. The system size is $n_x = n_y = 64$, $n_z = 128$ and the surfactant interaction strenghs are fixed at $g_{rs} = g_{bs} = g_{ss} = -0.005$. The capillary number is varied by changing the shear rate. The high capillary numbers are only reached by using the surface tension lowering effect of the added surfactant, hence the varying ranges of the datasets presented here. As the capillary number increases, $\theta_d$ can first increase beyond an angle of $45^\circ$ (dashed line) due to viscous effects. Only when the forces induced by the shear start to dominate does the droplet align with the shear flow~\cite{bib:singh-sarkar:2011}.}
\label{fig:plot-sheared-surfactant-inclination}
\end{figure}

For small capillary number, Taylor predicts a linear dependence of the deformation of a droplet on the capillary number~\cite{bib:taylor:1932,bib:taylor:1934}, with a particularly simple form for equiviscous fluids ($\lambda \equiv \mu_d / \mu_m = 1$):
\begin{equation}
  \label{eq:taylor-deformation}
  D = \frac{19 \mu_d + 16 \mu_m}{16 \mu_d + 16 \mu_m}  \mathrm{Ca} = \frac{35}{32} \mathrm{Ca}
\end{equation}
This law has been recovered in our simulations for the case of binary systems with various system sizes and interaction strengths, using the effective capillary number introduced in section~\ref{ssec:deformation-theory}.
Combining \eqnref{Ca-eff} and \eqnref{Re-eff} one can derive a relation between the capillary and Reynolds number:
\begin{equation}
  \label{eq:Re-Ca}
  \mathrm{Re}^{\mathrm{eff}} = \sigma \left( \frac{\rho_m R_d}{\mu_m^2} \right) \mathrm{Ca}^{\mathrm{eff}}
  \hbox{.}
\end{equation}
As we change the capillary number explicitly by changing the shear rate, the Reynolds number is proportional to the capillary number for a fixed value of the surface tension. Inertial effects increase the deformation, thus the deformations at high capillary number are higher than predicted by the linear relation of Taylor.

When a surfactant is added to the system, it lowers the surface tension of the interface, affecting the capillary number (but leaving the Reynolds number unchanged). Interaction strengths $g_{br} = 0.10$ and $g_{br} = 0.13$ are used, while the surfactant interaction strengths are fixed at $g_{rs} = g_{bs} = g_{ss} -0.005$ for the reasons mentioned in section~\ref{ssec:surfacetension-amphiphiles}. The initial homogeneous surfactant densities range from $\rho^s_{\mathrm{init}} = 0.0$ to $\rho^s_{\mathrm{init}} = 0.3$ in increments of $0.05$ and the system size is $n_x = n_y = 64$, $n_z = 128$, with an initial droplet radius of $R^{\mathrm{init}}_d = 0.3 \cdot n_x = 19.2$. The deformations for these systems are shown in \figref{plot-sheared-surfactant-deformation}. Since the change in surface tension directly enters the capillary number, all curves (including those not shown here for clarity) collapse onto a universal curve as a function of $\mathrm{Ca}^{\mathrm{eff}}$, and Taylor's law is reproduced for small capillary numbers $0 < \mathrm{Ca}^{\mathrm{eff}} < 0.06$. In the inset we show the relation between the capillary and Reynolds numbers. It is clear that these relations are linear, the slopes are proportional to the surface tension (which is changed both implicitly and explicitly), and agree with the values predicted by \eqnref{Re-Ca}.

Inclination angles of the droplet in its steady state are plotted in \figref{plot-sheared-surfactant-inclination}. In the case of Stokes flow, one would expect the inclination angle to be $45^\circ$ for very small capillary number and to observe a decrease of this angle as the capillary number is increased, indicating a better alignment of the droplet with the shear flow. However, as inertia plays a role here we observe that in some cases $\theta_d$ first increases beyond $45^\circ$, before the inclination decreases again and the droplet becomes elongated along the shear direction. When the steady inclinations are considered as a function of the Reynolds number, there exists a critical capillary number for which the inclination angle never exceeds $45^\circ$. Grouping the results at similar capillary numbers as datasets, we estimate this to be $\mathrm{Ca}^{\mathrm{eff}}_{\mathrm{crit}} \approx 0.11$. Our observations are consistent with results obtained by Singh and Sarkar, using a front-tracking finite-difference method~\cite{bib:singh-sarkar:2011}.

We now consider a system with nanoparticles as additives. The fluid-fluid interaction strength is held fixed at $g_{br} = 0.10$. As before, the particles have a radius of $r_p = 5.0$ and are neutrally wetting ($\theta_p = 90^\circ$). Initially, we choose their mass to be $m_p = 524$, as in section~\ref{ssec:surfactension-nanoparticles}. As discussed previously, the introduction of finite-sized particles introduces a lower bound on how small the simulation volume can be to accomodate enough particles on the interface and to avoid finite-size effects. For this reason, the simulation volume is chosen to be $n_x = n_y = 256$, $n_z = 512$, with an initial droplet radius of $R^{\mathrm{init}}_d = 0.3 \cdot n_x = 76.8$, still keeping it as small as possible to avoid excessive calculation time. The number of particles is varied as $n_p = 0$, $128$, $256$, $320$, $384$, $446$ and $512$, which results in a surface coverage fraction of $\chi = 0$ up to $\chi = 0.55$. Again, the capillary number is changed by changing the shear rate. Some examples of the deformations thus realised are shown in \figref{pictures-sheared-particles}, for $\mathrm{Ca}^{\mathrm{eff}} = 0.04$, $0.08$, $0.12$ and $\chi = 0.0$ (a), $\chi = 0.27$ (b) and $\chi = 0.55$ (c).

\begin{figure}
\includegraphics[width=\figwidth]{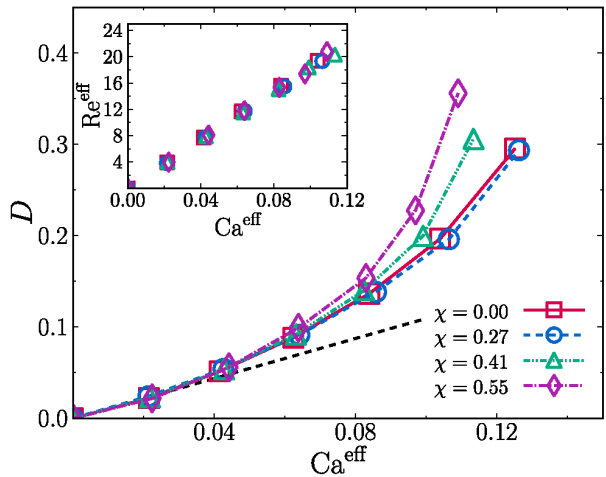}
\caption{Main plot: deformation parameter $D$ as a function of the effective capillary number $\mathrm{Ca}^{\mathrm{eff}}$ for various degrees of droplet interface particle coverage fraction $\chi$. In all these simulations $g_{br} = 0.10$, $r_p = 5.0$, $m_p = 524$, and $\theta_p = 90^\circ$. The system size is $n_x = n_y = 256$, $n_z = 512$. The capillary number is varied by changing the shear rate. Lines are added to guide the eye and clarify that the effect of adsorped particles is very weak for low $\chi$, but that the effect becomes noticeable at $\chi > 0.4$, where the deformation increases with $\chi$ at constant capillary number. In all cases, however, Taylor's law is reproduced for small $\mathrm{Ca}$ (dashed line). Note that the squares in this plot correspond to the triangles in \figref{plot-sheared-effective-Ca}. Inset: as is the case with surfactant, the Reynolds number scales linearly with the capillary number. Since the nanoparticles do not affect the surface tension, all curves have the same slope (cf. \eqnref{Re-Ca}).}
\label{fig:plot-sheared-particles-deformation}
\end{figure}

\begin{figure}
\includegraphics[width=\figwidth]{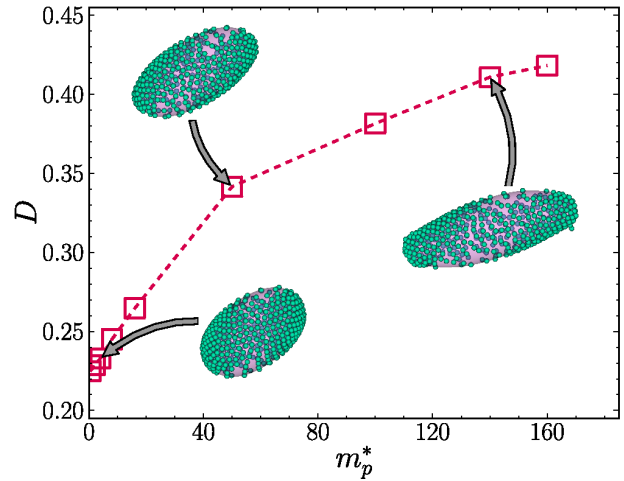}
\caption{The deformation parameter $D$ is shown as a function of the rescaled mass of the particles $m_p^* = m_p/m_p^0$, where $m_p^0 = 524$ is defined by setting the density of the particle to 1. The particles have a radius $r_p = 5.0$, their coverage fraction is $\chi = 0.55$ and the capillary number is $\mathrm{Ca}^{\mathrm{eff}} = 0.1$. Snapshots of the droplets are included, showcasing the deformations of the droplet. The inertia of the heavier particles causes additional deformation as they drag the droplet interface in the direction of the shear flow.}
\label{fig:plot-deformation-mass}
\end{figure}

Although the effect of addition of surfactant on the deformation and inclination of the droplet is automatically captured by the definition of the capillary number, the adsorped nanoparticles cause deviations from the previously observed behaviour. At low capillary number and low particle coverage, no differences are apparent and Taylor's law is reproduced (cf.~\figref{plot-sheared-particles-deformation}). When the coverage fraction grows beyond $\chi > 0.40$ the deformations in this regime increase with increasing $\chi$ and constant capillary number. As it was the case for the system with surfactant, the Reynolds number scales linearly with capillary number. However, since the nanoparticles do not affect the surface tension, all curves have the same slope (cf. inset of~\figref{plot-sheared-particles-deformation} and \eqnref{Re-Ca}). This implies that the increased deformation in the case of added nanoparticles is not caused by changes in inertia of the fluids. On the other hand, the inertia of the particles themselves plays a decisive role here. We have investigated the dependence of the droplet deformation on the size and mass of the particles. Particle radii have been varied between $4.0 \le r_p \le 10.0$ and at $\mathrm{Ca}^{\mathrm{eff}} = 0.1$ this has led to only a small change in $D$. Yet, changing the mass of the particles directly has a substantial effect. We have varied the mass of the particles over two orders of magnitude, as shown in \figref{plot-deformation-mass}. $\chi = 0.55$ and $\mathrm{Ca}^{\mathrm{eff}} = 0.1$ are kept constant and we have rescaled the mass scale with the reference mass: $m_p^* = m_p / 524$. The particles are accelerated as long as they are on the part of the droplet interface that experiences a shear flow at least partially parallel to the particle movement. Eventually, particles have to ``round the corner'' and are forced to move perpendicular to or even antiparallel to the shear flow. The increased inertia of heavier particles makes it more difficult to change the movement of these particles, leading to a situation where the droplet interface is in fact initially dragged farther away in the direction of the shear flow instead. This process is balanced by the surface tension as the surface area increases. This then explains the increase of deformation with increasing particle mass. As our deformation is increased substantially, the system size limits the deformation we can induce. Therefore, the values presented here are underpredictions of the actual effect of increased mass at high deformations, and might indeed hide a breakup event.

\begin{figure}
\includegraphics[width=\figwidth]{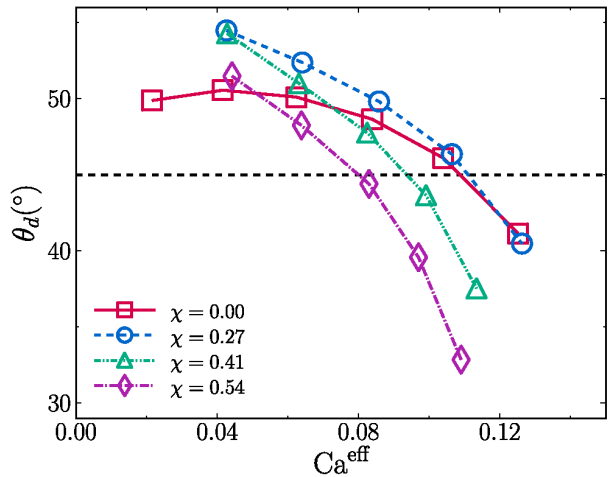}
\caption{Inclination angle $\theta_d$ as a function of the effective capillary number $\mathrm{Ca}^{\mathrm{eff}}$ for various degrees of droplet interface particle coverage fraction $\chi$. In all these simulations $g_{br} = 0.10$, $r_p = 5.0$, $m_p = 524$ and $\theta_p = 90^\circ$. The system size is $n_x = n_y = 256$, $n_z = 512$. The capillary number is varied by changing the shear rate. Due to the higher Reynolds numbers in these simulations when compared to the previous system, the inclination angle surpasses the $45^\circ$ mark (dashed line) in all cases, even in the case without particles. As in the study of deformation, the effect of a small number of particles seems to be relatively minor, but for $\chi > 0.4$, the inclination angle decreases sharply.}
\label{fig:plot-sheared-particles-inclination}
\end{figure}

The effect of particles on the inclination angle of the droplet is quantified in \figref{plot-sheared-particles-inclination}. We now return to using particles with mass $m_p = 524$. At low capillary number the disturbance caused to the droplet shape by the particles makes the inclination hard to measure. Due to the higher Reynolds numbers in these simulations when compared to the system with added surfactant, the inclination angle surpasses the $45^\circ$ mark in all cases, even in the case without particles at all~\cite{bib:singh-sarkar:2011}. As in the study of deformation, the effect of a small number of particles is relatively minor, but for $\chi > 0.4$, the inclination angle decreases sharply, as the droplet becomes more elongated and aligned with the shear flow. Increasing the particle mass also lowers the inclination angle, for the reasons described above, as can be observed in the droplet snapshots in \figref{plot-deformation-mass}.

\subsection{Droplet breakup}
\label{ssec:droplet-breakup}
\begin{figure}
\begin{tabular}{llll}
a) & & &\\
& \includegraphics[width=0.40\linewidth]{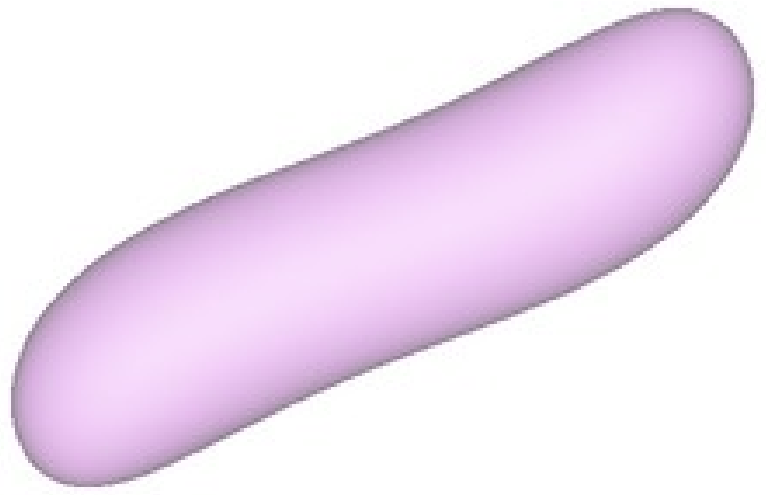} &
& \includegraphics[width=0.40\linewidth]{Fig17blank.eps} \\
\end{tabular}
\begin{tabular}{llll}
b)\vspace*{-0.3cm} & & c) & \\
& \includegraphics[width=0.40\linewidth]{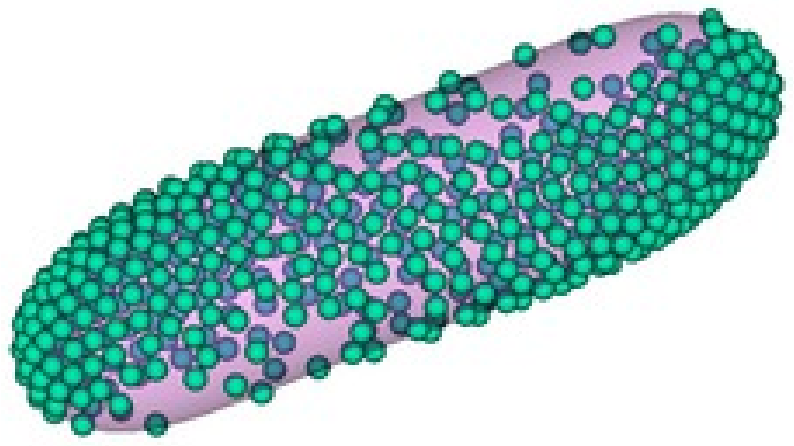} &
& \includegraphics[width=0.40\linewidth]{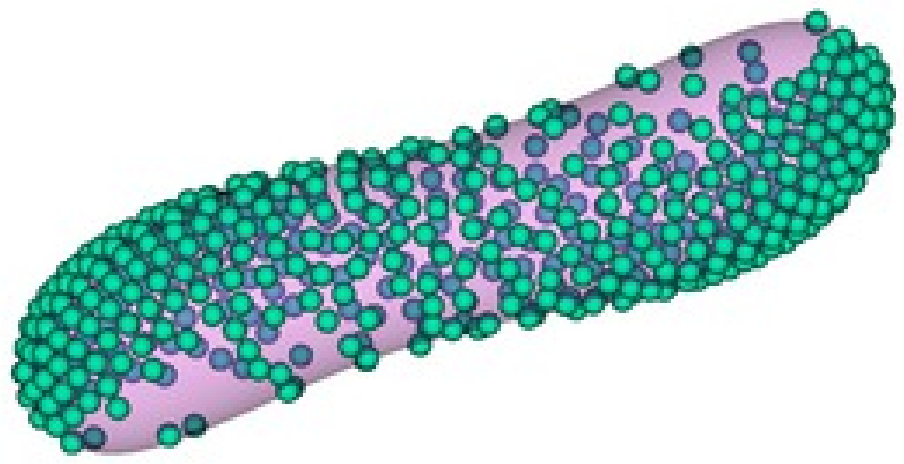} \\
d)\vspace*{-0.3cm} & & e) & \\
& \includegraphics[width=0.40\linewidth]{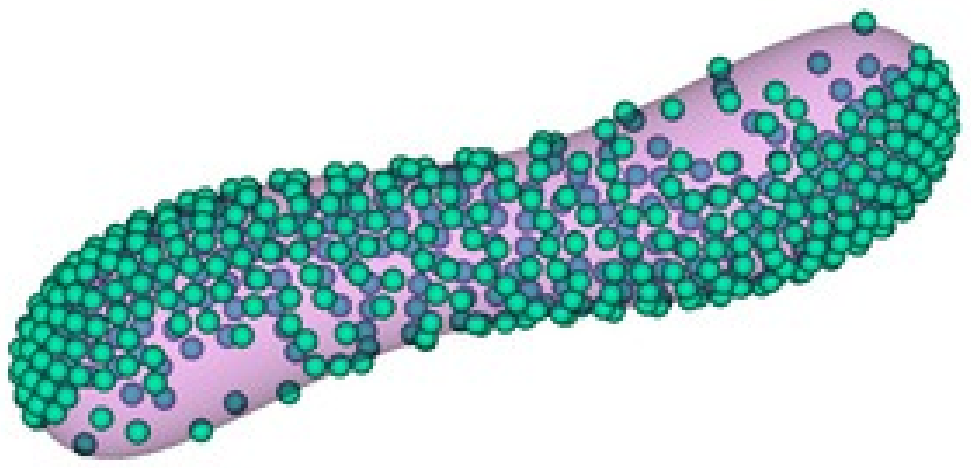} &
& \includegraphics[width=0.40\linewidth]{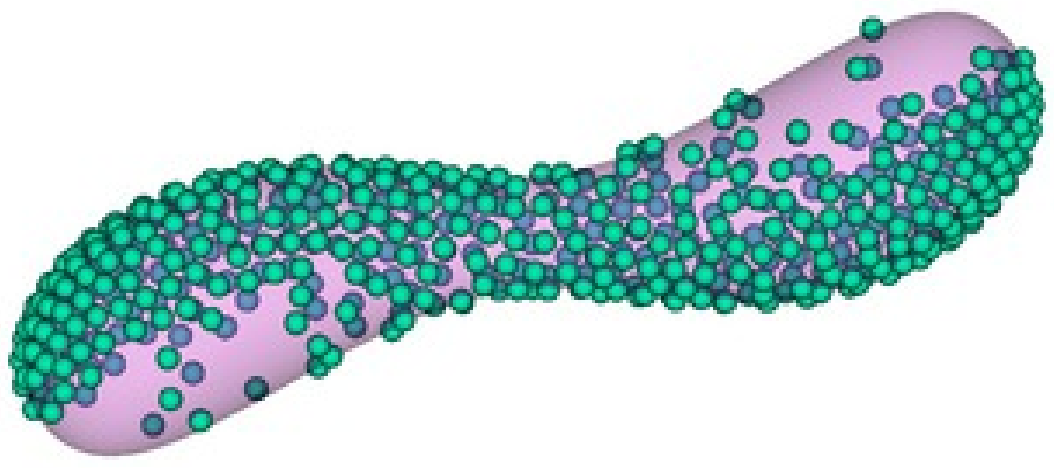} \\
f)\vspace*{-0.3cm} & & g) & \\
& \includegraphics[width=0.40\linewidth]{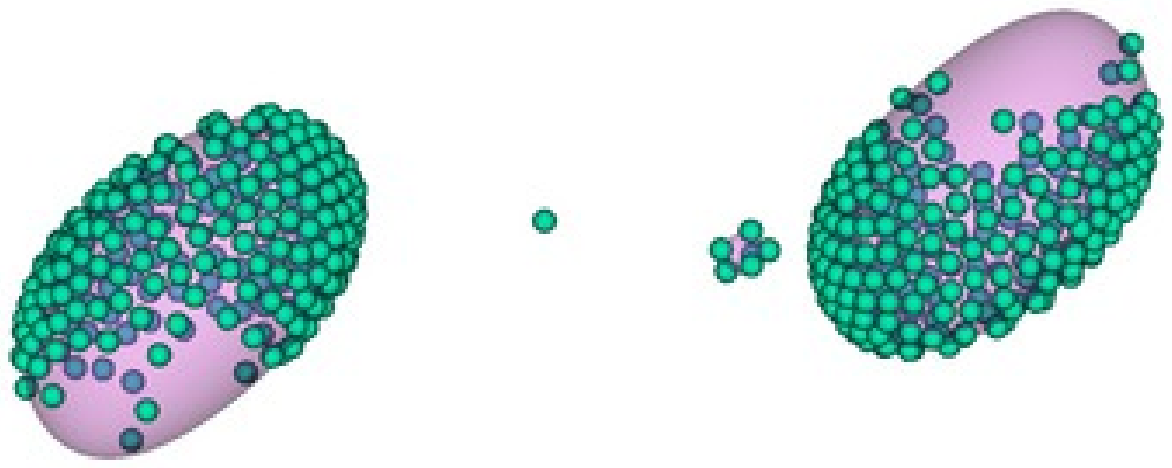} &
& \includegraphics[width=0.40\linewidth]{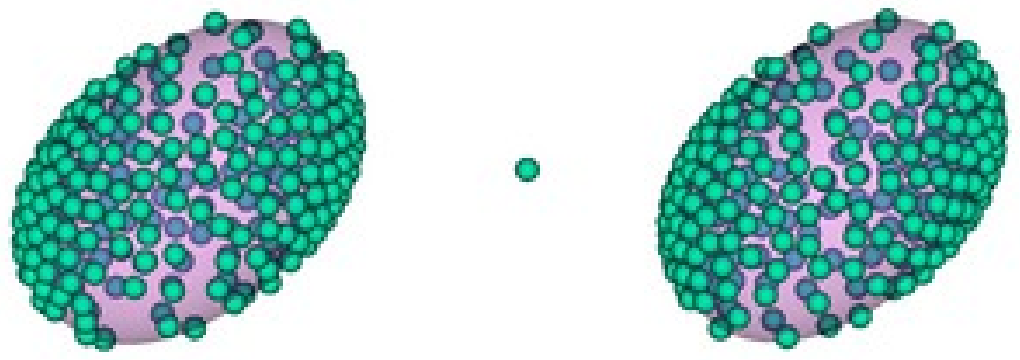} \\
\end{tabular}
\caption{Example of the breakup of a droplet when subjected to shear flow. The systems shown are identical ($n_x = n_y = 256$, $n_z = 512$, $g_{br} = 0.10$, $u_s = 0.07$, $R_d^{\mathrm{init}} = 76.8$, $\mathrm{Ca}^{\mathrm{eff}} = 0.15$) apart from the introduction of neutrally wetting particles of radius $r_p = 5.0$: a) $\chi = 0.0$, b--g) $\chi = 0.55$. In b-g) snapshots of the particle-covered droplet at different times are shown. The system without particles has reached a steady state at $t = 100000$ (a). At the same applied shear velocity, the droplet with the particles breaks up into two droplets of similar size. At b) $t = 50000$, c) $t = 70000$ d) $t = 80000$ and e) $t = 90000$ the droplet still holds together, but the deformation is extreme, departing from the ellipsoidal approximation and displaying a clear pinch-off. At f) $t = 100000$ the droplet has broken up into two similar-sized droplets, with the particles still distributed much as they were on the original droplet. After some relaxation the particles have redistributed themselves on the new interfaces at g) $t = 150000$.}
\label{fig:picture-breakup-nobreakup}
\end{figure}

When the capillary number is increased beyond the values shown in this work so far, we first proceed into a regime of extreme droplet deformation, where ellipsoidal approximations of the droplet shape no longer hold. This is followed by a regime of droplet breakup, where the surface tension cannot keep the droplet together and two or more smaller droplets form. Their increased relative surface area and smaller volume render them more stable against new deformations or breakup events. A series of snapshots of this process is shown in~\figref{picture-breakup-nobreakup}. First, a droplet without particles is shown in its steady state (a), strongly deformed at an applied shear velocity $u_s = 0.07$, but not breaking up. At the same applied shear velocity, a particle-covered droplet evolving in time is shown. First, deformations take place within the ellipsoidal approximation (b \& c). As the droplet is deformed even more, a definite neck is observed (d \& e). When this neck pinches off, two droplets are formed. In the highly deformed state just before breakup, the particles are mostly found near the centre of the $x$-direction, on the parts of the interface with highest curvature (this is an extreme example of the distributions described in section~\ref{ssec:distribution}). This means that just after the breakup, even though the new droplets are not very strongly deformed, there is a large anisotropy in the distribution of the particles, that is, one side of each droplet is mostly vacant (f). After more relaxation, however, the particles redistribute themselves over the interface in a similar fashion as before (g). Analyzing this behaviour in detail remains outside the scope of this work. We do remark that introducing adsorped particles decreases the resilience of the droplet against breakup, effectively lowering the critical capillary number at which breakup occurs, which can be viewed as an extension of the increased deformations.

\section{Conclusion} 
\label{sec:conclusion}

In this work we have applied our implementation of the lattice Boltzmann method, extended to deal with multiple fluid components, surfactants and hard-sphere nanoparticles to study physical phenomena related to a droplet in shear flow. Surface tensions in a binary system can be mapped to the choice of interaction strength between the fluid components and can be further adjusted by addition of a surfactant species. In this way, the surface tension can be varied by an order of magitude within the stable parameter region. The addition of spherical, neutrally wetting particles to the droplet interface does not affect the surface tension, owing to the fact that these only change interfacial free energy by removing part of the energetically unfavourable fluid-fluid interface.

When a droplet is subjected to simple shear flow, one of the characterizations of the system is the capillary number, relating the magnitude of the viscous forces to the magnitude of the surface tension. We have found that a measured effective shear rate better characterizes the system than the imposed shear rate, owing to the distortion in the velocity fields created by the presence of the droplet.

We have recovered Taylor's law for small deformations of a binary droplet, obtaining linear behaviour with the analytically predicted slope. For higher capillary number, the deformation increases more strongly than this linear relation. The surfactant model also conforms to this law: when surfactant is introduced into the system the capillary number is changed through the induced change in surface tension. Therefore, the same curve as found for the binary system is recovered. 

The effect of the addition of nanoparticles adsorped to the droplet interface on the deformation properties of the droplet has been studied. The particles are not homogeneously distributed over the droplet surface, but form more densely packed patches in areas with low shear velocities and high curvature. This pattern is in a dynamic equilibrium, and particles rotate over the droplet interface. Their rotational frequency increases with capillary number and decreases with distance from the centre of the system. For low capillary number or low coverage of the interface the effect of these nanoparticles is negligible. However, in the regime of high capillary number and high coverage ($\approx 50 \%$ in the undeformed state) the presence of particles induces a larger deformation at constant capillary number and a decrease in inclination angle. This is caused by the inertia of the massive particles. Finally, adsorped particles make the droplets break up more easily, lowering the critical capillary number at which breakup occurs. Emulsions consisting of such particle-covered droplets are expected to exhibit shear-thinning behaviour, as the increased deformation at higher shear rates lowers the apparent viscosity of such a complex fluid.

\begin{acknowledgments}
Financial support is acknowledged from the FOM/Shell IPP (09iPOG14 - ``Detection and guidance of nanoparticles for enhanced oil recovery'') and NWO/STW (Vidi grant 10787 of J.~Harting). We thank the J\"ulich Supercomputing Centre for the technical support and the CPU time which was allocated within a large scale grant of the Gauss Center for Supercomputing.
\end{acknowledgments}


\begin{thebibliography}{10}

\bibitem{bib:kim-stratford-cates:2010}
E.~Kim, K.~Stratford, and M.~Cates.
\newblock Bijels containing magnetic particles: A simulation study.
\newblock {\em Langmuir}, 26:7928, 2010.

\bibitem{bib:binks-fletcher:2001}
B.~Binks and P.~Fletcher.
\newblock Particles adsorped at the oil-water interface: A theoretical
  comparison between spheres of uniform wettability and ``{J}anus'' particles.
\newblock {\em Langmuir}, 17:4708, 2001.

\bibitem{bib:binks:2002}
B.~Binks.
\newblock Particles as surfactants -- similarities and differences.
\newblock {\em Cur. Opin. Colloid In.}, 7:21, 2002.

\bibitem{bib:tcholakova-denkov-lips:2008}
S.~Tcholakova, N.~Denkov, and A.~Lips.
\newblock Comparison of solid particles, globular proteins and surfactants as
  emulsifiers.
\newblock {\em Phys. Chem. Chem. Phys.}, 10:1608, 2008.

\bibitem{bib:gompper-schick:1994}
G.~Gompper and M.~Schick.
\newblock {\em Self-assembling amphiphilic systems}, volume~16.
\newblock Academic Press, 1994.

\bibitem{bib:chen-boghosian-coveney-nekovee:2000}
H.~Chen, B.~Boghosian, P.~Coveney, and M.~Nekovee.
\newblock A ternary lattice {B}oltzmann model for amphiphilic fluids.
\newblock {\em Proc. R. Soc. Lond. A}, 456:2043, 2000.

\bibitem{bib:harting-harvey-chin-venturoli-coveney:2005}
J.~Harting, M.~Harvey, J.~Chin, M.~Venturoli, and P.~V. Coveney.
\newblock Large-scale lattice {B}oltzmann simulations of complex fluids:
  advances through the advent of computational grids.
\newblock {\em Phil. Trans. R. Soc. Lond. A}, 363:1895, 2005.

\bibitem{bib:giupponi-harting-coveney:2006}
G.~Giupponi, J.~Harting, and P.~Coveney.
\newblock Emergence of rheological properties in lattice {B}oltzmann
  simulations of gyroid mesophases.
\newblock {\em Europhys. Lett.}, 73:533, 2006.

\bibitem{bib:ramsden:1903}
W.~Ramsden.
\newblock Separation of solids in the surface-layers of solutions and
  `suspensions'.
\newblock {\em Proc. R. Soc. Lond.}, 72:156, 1903.

\bibitem{bib:pickering:1907}
S.~Pickering.
\newblock Emulsions.
\newblock {\em J. Chem. Soc., Trans.}, 91:2001, 1907.

\bibitem{bib:arditty-whitby-binks-schmitt-lealcalderon:2003}
S.~Arditty, C.~Whitby, B.~Binks, V.~Schmitt, and F.~Leal-Calderon.
\newblock Some general features of limited coalescence in solid-stabilized
  emulsions.
\newblock {\em Eur. Phys. J. E}, 11:273, 2003.

\bibitem{bib:arditty-schmitt-giermannskakhan-lealcalderon:2004}
S.~Arditty, V.~Schmitt, J.~Giermannska-Kahn, and F.~Leal-Calderon.
\newblock Materials based on solid-stabilized emulsions.
\newblock {\em J. Colloid Interf. Sci.}, 275:659, 2004.

\bibitem{bib:binks-clint-whitby:2005}
B.~Binks, J.~Clint, and C.~Whitby.
\newblock Rheological behavior of water-in-oil emulsions stabilized by
  hydrophobic bentonite particles.
\newblock {\em Langmuir}, 21:5307, 2005.

\bibitem{bib:stratford-adhikari-pagonabarraga-desplat-cates:2005}
K.~Stratford, R.~Adhikari, I.~Pagonabarraga, J.-C. Desplat, and M.~Cates.
\newblock Colloidal jamming at interfaces: A route to fluid-bicontinuous gels.
\newblock {\em Science}, 309:2198, 2005.

\bibitem{bib:herzig-white-schofield-poon-clegg:2007}
E.~Herzig, K.~White, A.~Schofield, W.~Poon, and P.~Clegg.
\newblock Bicontinuous emulsions stabilized solely by colloidal particles.
\newblock {\em Nature Materials}, 6:966, 2007.

\bibitem{bib:clegg-herzig-schofield-egelhaalf-horozov-binks-cates-poon:2007}
P.~Clegg, E.~Herzig, A.~Schofield, S.~Egelhaaf, T.~Horozov, B.~Binks, M.~Cates,
  and W.~Poon.
\newblock Emulsification of partially miscible liquids using colloidal
  particles: Nonspherical and extended domain structures.
\newblock {\em Langmuir}, 23:5984, 2007.

\bibitem{bib:kim-stratford-adhikari-cates:2008}
E.~Kim, K.~Stratford, R.~Adhikari, and M.~Cates.
\newblock Arrest of fluid demixing by nanoparticles: A computer simulation
  study.
\newblock {\em Langmuir}, 24:6549, 2008.

\bibitem{bib:jansen-harting:2011}
F.~Jansen and J.~Harting.
\newblock From {B}ijels to {P}ickering emulsions: A lattice {B}oltzmann study.
\newblock {\em Phys. Rev. E}, 83:046707, 2011.

\bibitem{bib:aland-lowengrub-voigt:2011}
S.~Aland, J.~Lowengrub, and A.~Voigt.
\newblock A continuum model of colloid-stabilized interfaces.
\newblock {\em Phys. Fluids}, 23:062103, 2011.

\bibitem{bib:guenther-janoschek-frijters-harting:2012}
F.~G{\"u}nther, F.~Janoschek, S.~Frijters, and J.~Harting.
\newblock Lattice {B}oltzmann simulations of anisotropic particles at liquid
  interfaces.
\newblock {\em Comput. Fluids}, In press, 2012.
\newblock http://arxiv.org/abs/1109.3277.

\bibitem{bib:binks-horozov:2006}
B.~Binks and T.~Horozov.
\newblock {\em Colloidal Particles at Liquid Interfaces}.
\newblock Cambridge University Press, Cambridge, England, 2006.

\bibitem{bib:degraaf-dijkstra-vanroij:2010}
J.~de~Graaf, M.~Dijkstra, and R.~van Roij.
\newblock Adsorption trajectories and free-energy separatrices for colloidal
  particles in contact with a liquid-liquid interface.
\newblock {\em J. Chem. Phys.}, 132:164902, 2010.

\bibitem{bib:bleibel-dietrich-dominguez-oettel:2011}
J.~Bleibel, S.~Dietrich, A.~Dom{\'i}nguez, and M.~Oettel.
\newblock Shock waves in capillary collapse of colloids: A model system for
  two-dimensional screened newtonian gravity.
\newblock {\em Phys. Rev. Lett.}, 107:128302, 2011.

\bibitem{bib:bleibel-dominguez-oettel-dietrich:2011}
J.~Bleibel, A.~Dom{\'i}nguez, M.~Oettel, and S.~Dietrich.
\newblock Collective dynamics of colloids at fluid interfaces.
\newblock {\em Eur. Phys. J. E}, 34:125, 2011.

\bibitem{bib:janssen-vananroye-vanpuyvelde-moldenaers-anderson:2010}
P.~Janssen, A.~Vananroye, P.~van Puyvelde, P.~Moldenaers, and P.~Anderson.
\newblock Generalized behavior of the breakup of viscous drops in confinements.
\newblock {\em J. Rheol.}, 54:1047, 2010.

\bibitem{bib:succi:2001}
S.~Succi.
\newblock {\em The Lattice {B}oltzmann Equation for Fluid Dynamics and Beyond}.
\newblock Numerical Mathematics and Scientific Computation. Oxford University
  Press, Oxford, 2001.

\bibitem{bib:shan-chen:1993}
X.~Shan and H.~Chen.
\newblock Lattice {B}oltzmann model for simulating flows with multiple phases
  and components.
\newblock {\em Phys. Rev. E}, 47:1815, 1993.

\bibitem{bib:shan-chen:1994}
X.~Shan and H.~Chen.
\newblock Simulation of nonideal gases and liquid-gas phase transitions by the
  lattice {B}oltzmann equation.
\newblock {\em Phys. Rev. E}, 49:2941, 1994.

\bibitem{bib:orlandini-swift-yeomans:1995}
E.~Orlandini, M.~R. Swift, and J.~M. Yeomans.
\newblock A lattice {B}oltzmann model of binary-fluid mixtures.
\newblock {\em Europhys. Lett.}, 32:463, 1995.

\bibitem{bib:swift-orlandini-osborn-yeomans:1996}
M.~R. Swift, E.~Orlandini, W.~R. Osborn, and J.~M. Yeomans.
\newblock Lattice-{B}oltzmann simulations of liquid-gas and binary fluid
  systems.
\newblock {\em Phys. Rev. E}, 54:5041, 1996.

\bibitem{bib:dupin-halliday-care:2003}
M.~Dupin, I.~Halliday, and C.~Care.
\newblock Multi-component lattice {B}oltzmann equation for mesoscale blood
  flow.
\newblock {\em J. Phys. A: Math. Gen.}, 36:8517, 2003.

\bibitem{bib:lishchuk-care-halliday:2003}
S.~Lishchuk, C.~Care, and I.~Halliday.
\newblock Lattice {B}oltzmann algorithm for surface tension with greatly
  reduced microcurrents.
\newblock {\em Phys. Rev. E}, 67:036701, 2003.

\bibitem{bib:nekovee-coveney-chen-boghosian:2000}
M.~Nekovee, P.~Coveney, H.~Chen, and B.~Boghosian.
\newblock Lattice-{B}oltzmann model for interacting amphiphilic fluids.
\newblock {\em Phys. Rev. E}, 62:8282, 2000.

\bibitem{bib:ladd:1994:1}
A.~Ladd.
\newblock Numerical simulations of particulate suspensions via a discretized
  {B}oltzmann equation. {P}art {I}. {T}heoretical foundation.
\newblock {\em J. Fluid Mech.}, 271:285, 1994.

\bibitem{bib:ladd:1994:2}
A.~Ladd.
\newblock Numerical simulations of particulate suspensions via a discretized
  {B}oltzmann equation. {P}art {I}{I}. {N}umerical results.
\newblock {\em J. Fluid Mech.}, 271:311, 1994.

\bibitem{bib:ladd-verberg:2001}
A.~Ladd and R.~Verberg.
\newblock Lattice-{B}oltzmann simulations of particle-fluid suspensions.
\newblock {\em J. Stat. Phys.}, 104:1191, 2001.

\bibitem{bib:joshi-sun:2009}
A.~Joshi and Y.~Sun.
\newblock Multiphase lattice {B}oltzmann method for particle suspensions.
\newblock {\em Phys. Rev. E}, 79:066703, 2009.

\bibitem{bib:joshi-sun:2010}
A.~Joshi and Y.~Sun.
\newblock Wetting dynamics and particle deposition for an evaporating colloidal
  drop: A lattice {B}oltzmann study.
\newblock {\em Phys. Rev. E}, 82:041401, 2010.

\bibitem{bib:chen-doolen:1998}
S.~Chen and G.~Doolen.
\newblock Lattice {B}oltzmann method for fluid flows.
\newblock {\em Annu. Rev. Fluid Mech.}, 30:329, 1998.

\bibitem{bib:sukop-thorne:2007}
M.~Sukop and D.~Thorne.
\newblock {\em Lattice {B}oltzmann Modelling - An introduction for
  geoscientists and engineers}.
\newblock Springer Berlin Heidelberg, 2007.

\bibitem{bib:bhatnagar-gross-krook:1954}
P.~Bhatnagar, E.~Gross, and M.~Krook.
\newblock A model for collision processes in gases. {I}. {S}mall amplitude
  processes in charged and neutral {one-component} systems.
\newblock {\em Phys. Rev. E}, 94:511, 1954.

\bibitem{bib:chen-chen-matthaeus:1992}
H.~Chen, S.~Chen, and W.~Matthaeus.
\newblock Recovery of the {N}avier-{S}tokes equations using a lattice-gas
  {B}oltzmann method.
\newblock {\em Phys. Rev. A}, 45:R5339, 1992.

\bibitem{bib:benzi-chibbaro-succi:2009}
R.~Benzi, S.~Chibbaro, and S.~Succi.
\newblock Mesoscopic lattice {B}oltzmann modeling of flowing soft systems.
\newblock {\em Phys. Rev. Lett.}, 102:026002, 2009.

\bibitem{bib:benzi-sbragaglia-succi-bernaschi-chibbaro:2009}
R.~Benzi, M.~Sbragaglia, S.~Succi, M.~Bernaschi, and S.~Chibbaro.
\newblock Mesoscopic lattice {B}oltzmann modeling of soft-glassy systems:
  Theory and simulations.
\newblock {\em J. Chem. Phys.}, 131:104903, 2009.

\bibitem{bib:furtado-skartlien:2010}
K.~Furtado and R.~Skartlien.
\newblock Derivation and thermodynamics of a lattice {B}oltzmann model with
  soluble amphiphilic surfactant.
\newblock {\em Phys. Rev. E}, 81:066704, 2010.

\bibitem{bib:aidun-lu-ding:1998}
C.~Aidun, Y.~Lu, and E.-J. Ding.
\newblock Direct analysis of particulate suspensions with inertia using the
  discrete {B}oltzmann equation.
\newblock {\em J. Fluid Mech.}, 373:287, 1998.

\bibitem{bib:hertz:1881}
H.~Hertz.
\newblock {\"U}ber die {B}er\"uhrung fester elastischer {K}\"orper.
\newblock {\em Journal f\"ur die reine und angewandte Mathematik}, 92:156,
  1881.

\bibitem{bib:stratford-adhikari-pagonabarraga-desplat:2005}
K.~Stratford, R.~Adhikari, I.~Pagonabarraga, and J.-C. Desplat.
\newblock Lattice {B}oltzmann for binary fluids with suspended colloids.
\newblock {\em J. Stat. Phys.}, 121:163, 2005.

\bibitem{bib:lees-edwards:1972}
A.~Lees and S.~Edwards.
\newblock The computer study of transport processes under extreme conditions.
\newblock {\em J. Phys. C.}, 5:1921, 1972.

\bibitem{bib:wagner-pagonabarraga:2002}
A.~Wagner and I.~Pagonabarraga.
\newblock Lees–{E}dwards boundary conditions for lattice {B}oltzmann.
\newblock {\em J. Stat. Phys.}, 107:521, 2002.

\bibitem{bib:schmieschek-harting:2011}
S.~Schmieschek and J.~Harting.
\newblock Contact angle determination in multicomponent lattice {B}oltzmann
  simulations.
\newblock {\em Commun. Comput. Phys.}, 9:1165, 2011.

\bibitem{bib:bauer:2000}
R.~Bauer.
\newblock Distribution of points on a sphere with application to star catalogs.
\newblock {\em J. Guid. Control Dynam.}, 23:130, 2000.

\bibitem{bib:taylor:1932}
G.~Taylor.
\newblock The viscosity of a fluid containing small drops of another fluid.
\newblock {\em Proc. R. Soc. Lond. A}, 138:41, 1932.

\bibitem{bib:taylor:1934}
G.~Taylor.
\newblock The formation of emulsions in definable fields of flow.
\newblock {\em Proc. R. Soc. Lond. A}, 146:501, 1934.

\bibitem{bib:stone-leal:1990}
H.~Stone and L.~Leal.
\newblock The effects of surfactants on drop deformation and breakup.
\newblock {\em J. Fluid Mech.}, 220:161, 1990.

\bibitem{bib:kaoui-harting-misbah:2011}
B.~Kaoui, J.~Harting, and C.~Misbah.
\newblock Two-dimensional vesicle dynamics under shear flow: effect of
  confinement.
\newblock {\em Phys. Rev. E}, 83:066319, 2011.

\bibitem{bib:singh-sarkar:2011}
R.~Singh and K.~Sarkar.
\newblock Inertial effects on the dynamics, streamline topology and interfacial
  stresses due to a drop in shear.
\newblock {\em J. Fluid Mech.}, 683:149, 2011.

\end{thebibliography}
\end{document}